\documentclass[structabstract]{aa}  
\usepackage{graphicx}
\usepackage{txfonts}
\usepackage{natbib}
\usepackage{color}

\begin{document}

\title{The HI/OH/Recombination line survey of the inner Milky Way (THOR) \thanks{Based on observations carried out with the Karl Jansky Very Large Array (VLA).}}

\subtitle{Survey overview and data release 1}

   \author{H.~Beuther
          \inst{1}
          \and
          S.~Bihr
          \inst{1}
          \and
          M.~Rugel
          \inst{1}
          \and
          K.~Johnston
          \inst{1,2}
          \and
          Y.~Wang
          \inst{1}
          \and
          F.~Walter
           \inst{1}
          \and
          A.~Brunthaler
           \inst{3}
          \and
          A.J.~Walsh
          \inst{4}
          \and
          J.~Ott
           \inst{5}
          \and
          J.~Stil
           \inst{6}
          \and
          Th.~Henning
           \inst{1}
          \and
          T.~Schierhuber
           \inst{1}
           \and
          J.~Kainulainen
           \inst{1}
           \and
          M.~Heyer
           \inst{7}
           \and
          P.F.~Goldsmith
           \inst{8}
           \and
          L.D.~Anderson
           \inst{9}
           \and
          S.N.~Longmore
           \inst{10}
           \and
          R.S.~Klessen
           \inst{11}
           \and
          S.C.O.~Glover
           \inst{11}
           \and
          J.S.~Urquhart
           \inst{3,17}
           \and
          R.~Plume
           \inst{6}
           \and
          S.E.~Ragan
           \inst{2}
           \and
          N.~Schneider
           \inst{12}
           \and
          N.M.~McClure-Griffiths 
           \inst{13}
           \and
          K.M.~Menten
           \inst{3}
           \and
          R.~Smith
           \inst{14}
           \and
          N.~Roy
           \inst{15}
          \and
          R.~Shanahan
           \inst{6}
          \and
          Q.~Nguyen-Luong
           \inst{16}
          \and
          F.~Bigiel
           \inst{11}
            }
   \institute{$^1$ Max Planck Institute for Astronomy, K\"onigstuhl 17,
              69117 Heidelberg, Germany, \email{beuther@mpia.de}\\
             $^2$ School of Physics and Astronomy, University of Leeds, Leeds, LS2 9JT, UK\\
             $^3$ Max Planck Institute for Radioastronomy, Auf dem H\"ugel 69, 53121 Bonn, Germany\\
             $^4$ International Centre for Radio Astronomy Research, Curtin University, GPO Box U1987, Perth WA 6845, Australia\\
             $^5$ National Radio Astronomy Observatory, P.O. Box O, 1003 Lopezville Road, Socorro, NM 87801, USA\\
             $^6$ Department of Physics and Astronomy, University of Calgary, 2500 University Drive NW, Calgary AB, T2N 1N4, Canada\\
             $^7$ Department of Astronomy, University of Massachusetts, Amherst, MA 01003-9305, USA\\
             $^8$ Jet Propulsion Laboratory, California Institute of Technology,
4800 Oak Grove Drive, Pasadena, CA 91109, USA\\
             $^9$ Department of Physics and Astronomy, West Virginia University, Morgantown, WV 26506, USA\\
             $^{10}$ Astrophysics Research Institute, Liverpool John Moores University, 146 Brownlow Hill, Liverpool L3 5RF, UK\\
             $^{11}$ Universit\"at Heidelberg, Zentrum f\"ur Astronomie, Institut f\"ur Theoretische Astrophysik, Albert-Ueberle-Str. 2, D-69120 Heidelberg, Germany\\
             $^{12}$ I. Physikalisches Institut, University of Cologne, Z\"ulpicher Str.~77, 50937 K\"oln, Germany\\
             $^{13}$ Research School of Astronomy and Astrophysics, The Australian National University, Canberra, ACT, Australia\\
             $^{14}$ Jodrell Bank Centre for Astrophysics, School of Physics and Astronomy, The University of Manchester, Oxford Road, Manchester,M13 9PL, UK\\
             $^{15}$ Department of Physics, Indian Institute of Science, Bangalore 560012, India\\
             $^{16}$ National Astronomical Observatory of Japan, Chile Observatory, 2-21-1 Osawa, Mitaka, Tokyo 181-8588, Japan\\
             $^{17}$ School of Physical Sciences, University of Kent, Ingram Building, Canterbury, Kent CT2 7NH, UK
}

   \date{Version of \today}

%   \abstract{}
% \abstract{}{}{}{}{} 
% 5 {} token are mandatory
\abstract
{The past decade has witnessed a large number of
  Galactic plane surveys at angular resolutions below
  $20''$. However, no comparable high-resolution survey exists at long
  radio wavelengths around 21\,cm in line and continuum emission.}
{We remedy this situation by studying the northern Galactic plane at
  $\sim$20$''$ resolution in emission of atomic, molecular, and ionized gas.}
{Employing the Karl G.~Jansky Very Large Array (VLA) in the C-array
  configuration and a large program, we observe the HI 21\,cm line,
  four OH lines, nineteen H$n\alpha$ radio recombination lines as well
  as the continuum emission from 1 to 2\,GHz in full polarization over
  a large part of the first Galactic quadrant.}
{Covering Galactic longitudes from 14.5 to 67.4\,deg and latitudes
  between $\pm 1.25$\,deg, we image all of these lines
  and the continuum at $\sim$20$''$ resolution. These data allow us to
  study the various components of the interstellar medium (ISM): from
  the atomic phase, traced by the HI line, to the molecular phase,
  observed by the OH transitions, to the ionized medium, revealed by
  the cm continuum and the H$n\alpha$ radio recombination
  lines. Furthermore, the polarized continuum emission enables
  magnetic field studies. In this overview paper, we discuss the
  survey outline and present the first data release as well as early
  results from the different datasets. We now release the first half
  of the survey; the second half will follow later after the ongoing
  data processing has been completed. The data in fits format
  (continuum images and line data cubes) can be accessed through the
  project web-page http://www.mpia.de/thor.}
{The HI/OH/Recombination line survey of the Milky Way (THOR){} opens a new
window to the different parts of the ISM. It enables detailed studies
of molecular cloud formation, conversion of atomic to molecular gas,
and feedback from H{\sc ii} regions as well as the magnetic field in the
Milky Way. It is highly complementary to other surveys of our
Galaxy, and comparing the different datasets will allow us to address
many open questions.}  \keywords{Stars: formation -- ISM:
    clouds -- ISM: structure -- ISM: kinematics and dynamic -- ISM: magnetic fields -- stars: evolution}

\titlerunning{THOR}

\maketitle

\section{Introduction}
\label{intro}

Over the past decade, the Galactic plane was surveyed comprehensively
from near-infrared to cm wavelengths.  These surveys enable
investigations of not only individual local phenomena such as stars,
clusters, ionized gas and molecular or atomic clouds, but studies
of our Galaxy as a whole, and we can compare the results to
extragalactic studies (see, e.g.,
\citealt{taylor2003,churchwell2009,carey2009,schuller2009,anderson2011,walsh2011,beuther2012b,ragan2014,wang2015,goodman2014,reid2014,abreu2016}).
Particularly important for a general understanding of the different
physical processes is the multiwavelength approach because different
surveys trace different components of the interstellar medium (ISM)
and stellar populations, as well as varying temperature regimes and
physical processes. Earlier ideas for such a multiwavelength
  survey approach were promoted by the Canadian Galactic Plane
  Survey \citep{taylor2003}, for example. The different phases
(atomic, molecular, or ionized gas and dust) are not isolated, but
interact and, maybe even more importantly, they  change from one
phase to the other in the natural matter cycle of the ISM.  It
is therefore important for our understanding of ISM dynamics and star
formation to have surveys at comparable angular resolution.

While most of the infrared to mm Galactic plane surveys have an
angular resolution better than $20''$, the existing HI Very Large
Array Galactic Plane Survey (VGPS) survey conducted with the Very
Large Array (VLA) in its compact D-configuration has an angular
resolution of only $60''$ \citep{stil2006}. For comparison, the most
recent single-dish survey of HI with the Effelsberg telescope has an
angular resolution of $10'$ \citep{winkel2016}. Even though the VLA
D-configuration as well as single-dish HI surveys are appropriate for
studying atomic Galactic structure on large scales, they are less
useful for the direct comparison with the other existing surveys
mentioned above. For example, previous 60$''$ resolution observations
of HI and CO emissions lines showed that large-scale atomic gas
envelopes and atomic gas flows in the surrounding environments are
needed to form denser molecular gas, and subsequently dense core and
massive stars (e.g., \citealt{nguyen2011,motte2014}). However, these
data could not yet be used to study the interaction between the atomic
and the dense molecular gas structures that may occur on significantly
smaller scales (see, e.g., the recent 870\,$\mu$m dust continuum
emission Galactic plane survey ATLASGAL at $19''$ resolution,
\citealt{schuller2009}). For reference, we mention that 0.5\,pc
corresponds to $25''$ at a typical molecular cloud distance of 4\,kpc.

Furthermore, the new capabilities of the WIDAR correlator at the VLA
allow us to observe many spectral lines simultaneously, in particular
several molecular OH transitions, a series of H$n\alpha$ radio
recombination lines (RRLs, $n=151 \ldots 186$), and the
continuum emission. Combining these data with the HI observations
probes the transition of matter in the ISM from the diffuse neutral
atomic to the dense molecular and the ionized gas components and
back. This combined approach is followed in The
HI/OH/Recombination line survey of the Milky Way (THOR) we present
here.  These new
THOR C-configuration HI data ($15''-20''$ resolution corresponding to
linear scales of 0.2-0.3\,pc at typical distances of 3\,kpc), when
combined with the existing D-configuration and GBT (Green Bank
Telescope) observations to include the larger-scale emission
\citep{stil2006}, enable us to address many questions associated with
atomic hydrogen from large-scale Galactic structure and cloud
formation processes down to the scales of individual star-forming
regions. At the same time, the OH, RRLs, and continuum data provide a
more complete picture of the Galactic ISM.

This paper presents an overview of THOR and the first data release. The
motivation and goals of the survey are described in Sect. 2, and the
parameters of the survey are presented in Sect. 3. The observation
details and data analysis are given in Sect. 4, while initial
results from this survey are presented in Sect. 5. Finally, the
potential of this survey and the future possibilities are discussed in
Sect. 6, and a summary is presented in Sect. 7.

\section{Goals of the survey}

\subsection{Atomic to molecular hydrogen transition of clouds}

Several cloud formation scenarios favor converging flows in which
large-scale gas streams collide and form density enhancements in which
the conversion from atomic to molecular hydrogen is thought to mainly take
place (e.g.,
\citealt{ballesteros1999,hartmann2001,vazquez2006,hennebelle2007,hennebelle2008,heitsch2008,banerjee2009,ballesteros2011,clark2012,dobbs2014}).
These simulations predict that before molecular gas forms, the
medium remains in an atomic phase for several million years (e.g.,
\citealt{elmegreen2007,heitsch2008b,clark2012}). Hydrodynamical
simulations coupled with chemical networks and radiative transfer
calculations provide predictions of spectral line parameters (e.g.,
line widths, spatially resolved kinematics) and physical and
structural
properties (e.g., probability density functions) of the different
phases of the ISM.

To investigate current cloud formation models, a sensitive
characterization of the atomic HI phase at an angular
resolution comparable to the molecular gas is therefore mandatory.  HI absorption line
studies at high angular and spectral resolution (FWHM of the
cold neutral
medium (CNM) in HI at 100\,K $\sim $2.2\,km\,s$^{-1}$) have proven to
be an excellent tool for studying the CNM and its association with dense
molecular gas cores (e.g.,
\citealt{heiles2003,dickey2003,li2003,goldsmith2005,gibson2005,krco2008,kanekar2011,roy2013,roy2013b,liszt2014,lee2015,murray2015}).
Combining the HI absorption lines with the simultaneously observed
molecular OH absorption or emission as well as other tracers of the cold,
dense ISM such as the submm continuum (e.g., the ATLASGAL survey,
\citealt{schuller2009}) allow studying the interplay and accretion of
the atomic and molecular gas from the larger cloud-scales
($\sim$10\,pc) to the smaller core-scales (0.25-0.5\,pc) (e.g.,
\citealt{goldbaum2011,smith2012}). Considering the sensitivity to both
small- and large-scale HI emission, this survey will also be very
useful for studying the second-order statistics, for instance, the angular
correlation function or power spectrum, of the cold atomic ISM, and
thus probe turbulence at these scales.

Complementary molecular gas information is available through several
surveys such as  the CO emission from the JCMT CO survey
\citep{dempsey2013,rigby2016}, the Galactic Ring Survey \citep{jackson2006}, the
Exeter-FCRAO CO Galactic Plane survey (Brunt et al.~in prep.), and
dense gas studies through spectroscopic follow-ups of the ATLASGAL and
BGPS (sub)mm continuum surveys (e.g.,
\citealt{schlingman2011,wienen2012,shirley2013,giannetti2014}).

THOR facilitates characterizing the phase transition from
atomic to molecular gas in detail, directly linking models with observations,
and comparing the results with extragalactic studies (e.g.,
\citealt{hennebelle2007b,glover2010,glover2011,shetty2011,walter2008,leroy2008,smith2014b,walch2015,girichidis2016,bertram2016}).

\subsection{OH maser and thermal emission and absorption} 

We conduct simultaneous observations of four hydroxyl lines.  Hydroxyl
(OH) has a remarkably constant abundance relative to molecular
hydrogen in diffuse and translucent molecular clouds up to column
densities of $\sim 10^{22}~\rm{cm}^{-2}$ \citep{lucas1996}. These
observations yield sensitive information on ISM gas with properties in
between cold neutral atomic and dense molecular gas that so far has
only been sparsely studied. Combining the information from the four
hyperfine structure lines at 1612, 1665, 1667, and 1720\,MHz (relative
intensities 1:5:9:1) can constrain the OH excitation behavior and
deliver local thermodynamic equilibrium (LTE) estimates of column
densities, as well as kinematics and neutral particle and electron
densities \citep{nguyen1976,elitzur1976,guibert1978}.

In addition to the thermal OH emission and absorption, we can identify a
flux-limited sample of OH masers at 1612, 1665, 1667, and 1720\,MHz in
the northern Milky Way. While previous surveys typically covered only
one line, observing all four maser lines is particularly interesting
since they trace different physical and evolutionary phases (see
Sect. \ref{oh}). In combination with the southern hemisphere single-dish OH
survey SPLASH \citep{dawson2014}, these observations will show the
full population of OH masers in the Galaxy.

\begin{figure*}[htb] 
\includegraphics[width=0.99\textwidth]{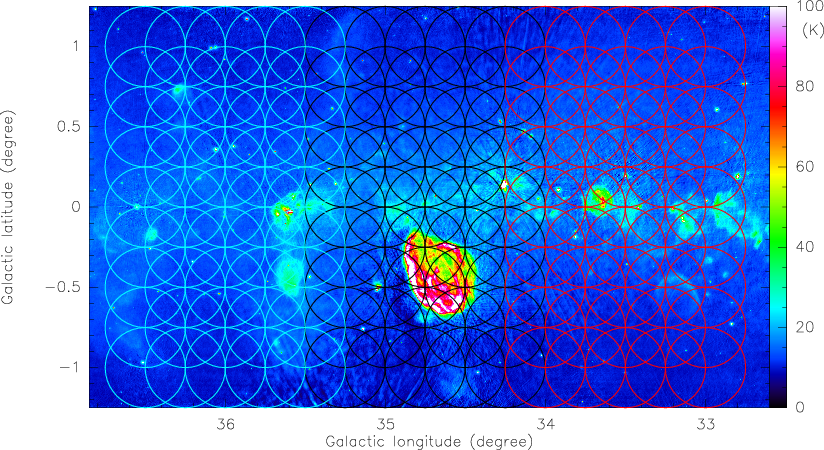}
\caption{Circles show the 45 mosaic pointings for three adjacent
  tiles centered on 33.5, 34.75, and 36.0 degrees Galactic longitude in
  red, black, and cyan, respectively. The circle diameters of $30'$
  correspond to the primary beam at 1.5\,GHz, and the pointings are
  separated by $15'$. The underlying image presents the THOR+VGPS
  continuum data at 1.4\,GHz converted into a temperature scale in K
  using the Rayleigh-Jeans approximation.}
\label{mosaic}
\end{figure*}

\subsection{H$n\alpha$ radio recombination lines}

Radio recombination lines trace the ionized gas of H{\sc ii}
regions. We are able to study its kinematics in a spatially resolved
fashion.  Combining the emission from the ionized gas with that of the
atomic (HI) and molecular components (e.g., JCMT CO survey,
\citealt{dempsey2013}; Galactic Ring Survey, \citealt{jackson2006}; or
dense gas surveys, \citealt{schlingman2011,wienen2012,shirley2013}) and the stellar components observable by Spitzer allows us to
study the expansion of ionized regions in the surrounding medium and
the (non-)association with young embedded stellar populations.  Such a
dataset enables investigating triggering processes in star formation
as well as general feedback processes of the different phases of the
ISM.

\subsection{Continuum emission}

Observing the continuum emission from 1 to 2\,GHz facilitates the
derivation of spectral indices for thousands of sources ranging from
H{\sc ii} regions to background galaxies. With these data we can
characterize the physical properties of the gas (e.g., electron number
densities), differentiate free-free from synchrotron emission, and
determine whether the gas is optically thin or thick. These data may
also be used to derive the Galactic continuum emission fluctuation
power spectrum of the diffuse component, which is related to the
density and magnetic field fluctuations (e.g.,
\citealt{goldreich1995,iacobelli2013}). Characterizing the low-frequency continuum is interesting in itself in addition to being useful for building up foreground models for ongoing and future low-frequency observations targeting cosmological signals. Furthermore, we
can resolve the kinematic distance ambiguity for H{\sc II} regions
using HI absorption against the H{\sc ii} region broadband continuum
emission (e.g., \citealt{kolpak2003,anderson2009}). Compared to the
VGPS, we have better sensitivity to small-scale structure, which
permits this analysis for fainter H{\sc ii} regions in more
complicated zones of star formation. While the Multi-Array
Galactic Plane Imaging Survey (MAGPIS, \citealt{helfand2006}) achieves an even
higher angular resolution, THOR is the only available L-band continuum
survey that allows us to derive the spectral index information.

\subsection{Polarization and magnetic field measurements}

Obtaining the linear polarization information for the continuum
facilitates the identification of distant young supernova remnants. In
addition, the Faraday effect, describing the rotation of the plane of
polarization that is due to magnetic fields along the line of sight,
can be used to determine magnetic field properties on large and small
scales toward polarized Galactic and extragalactic sources. The data
have higher angular resolution and L-band frequency coverage than
existing surveys. The higher angular resolution ($\sim$1\,pc at a
distance of 10\,kpc) will allow us to identify young supernova
remnants in crowded star formation regions where surveys like the NRAO
VLA Sky Survey (NVSS, \citealt{condon1998}) suffer from confusion with
bright thermal emission. Moreover, a smaller beam size reduces
depolarization of resolved sources by differential Faraday rotation
across the beam.

Integrating the polarized source counts of \citet{hales2014}, who had
similar resolution at 1.4\,GHz, we estimate that $\sim$490
extragalactic sources with a polarized flux density greater than 3\,mJy
($\sim 10\sigma$) exist in the $\sim$132 square degree survey
area. This may be compared with the 194 sources measured by
\citet{vaneck2011}, who targeted polarized sources selected from the
NVSS. The difference in sample size results from a combination of
factors that include bandwidth depolarization and confusion with
bright emission in the NVSS, a larger portion of resolved sources at
higher angular resolution, and specific target selection criteria
adopted by \citet{vaneck2011}. THOR will increase the sample of
extragalactic sources at very low Galactic latitude that probe the
entire Milky Way disk.

The intrinsic polarization of resolved supernova remnants provides
information on the magnetic field structure and the degree of order in the
magnetic field. The observed polarization is affected by Faraday
rotation by the (turbulent) foreground, and possibly internal Faraday
rotation. Beyond traditional measurements of Faraday rotation that
yield a single rotation measure, the wide frequency coverage of the
data allows us to investigate higher order effects that occur when
different parts of the source experience different amounts of Faraday
rotation (e.g., \citealt{farnsworth2011}). For example, differential
Faraday rotation across the synthesized beam by a turbulent plasma in
the line of sight to a resolved source results in wavelength-dependent
depolarization that can be detected in broad-band polarimetric
data. Modeling these effects will help to reconstruct the intrinsic
polarization of the supernova remnant, and provides information about
the turbulent medium in which the Faraday rotation occurs.

\section{Survey}
\label{survey}

This THOR 21\,cm line and continuum survey is a Large Program at the
Karl G.~Jansky Very Large Array with approximately 215\,hours of
observing time in the C-array configuration \citep{perley2011}. The
primary beam size over the L-band from 1 to 2\,GHz changes by a factor
2, and the absolute areal coverage of THOR depends slightly on the
spectral window considered.  The approximate areal coverage of THOR is
$\sim$132 square degrees from 14.5 to 67.25 degrees in Galactic
longitude and $\pm 1.25$\,degrees in Galactic latitude. This coverage
is based on scientific as well as technical arguments: Scientifically,
this part of the Milky Way covers a large portion of the inner Milky
Way that also includes the bar-spiral interface. Hence, very different
star formation environments are observable from very active almost
starburst-like regions (e.g., W43) out to less active environments in
the larger longitude range. From a technical point of view, this is
approximately the same coverage as the previous HI and 1.4\,GHz
continuum survey VGPS. This enables us to combine the new THOR
observations with the VGPS data to recover signal over a wider range
of spatial scales.  The survey as a whole was conducted during three
campaigns. It started with a pilot study in 2012 that targeted mainly
four square degrees around the mini-starburst W43 (legacy id
AB1409). Based on the very positive initial results from this pilot
study (e.g., \citealt{bihr2015b,walsh2016}), we conducted the first
half of the survey in 2013 (phase 1, id AB1447). It covered the
longitude range between 14.5 and 37.9 degrees and a smaller
strip from 47.1 to 51.2 degrees targeting the Sagittarius arm
tangential point, including the star-forming region W51. The combined
continuum data of THOR in the VLA C-configuration with previous VGPS
data (VLA D-configuration plus single-dish GBT observations,
\citealt{stil2006}) of this first half of the survey are shown in
Figs.~\ref{continuum1} and \ref{continuum2}. The second half of the
survey was observed from the end of 2014 to the beginning of 2015
(phase 2, id AB1513). While we report in this paper the survey,
calibration and imaging strategies of the full survey, we present here only
images and early results from the first half of the survey because
of the time-consuming nature of the imaging process. The
remaining data products will then be published in the near future.

We require an angular resolution of $\leq 20''$ so that the data are
comparable to other Galactic plane surveys (e.g., ATLASGAL,
\citealt{schuller2009}). This goal requires in the L-band (between 1
and 2\,GHz frequency) observations in the VLA C-configuration. To image the
most extended component, the atomic HI, and also for the 1.4\,GHz
continuum, these data can be combined with the previous VGPS survey
data \citep{stil2006}. For the remaining data, we concentrate on
observations in the C-configuration alone.

The new WIDAR correlator is extremely flexible, facilitating a broad
coverage of frequencies as well as zooms into many bands. The spectral
setup is designed with three main goals in mind: (a) to spectrally
resolve the HI and OH lines at comparable resolution
($\sim$1.5\,km\,s$^{-1}$) as the previous VGPS survey did for the HI
line. This is a good compromise between signal-to-noise ratio and
spectral resolving power for the thermal HI lines. (b) The second goal
is to observe the full L-band bandpass from 1 to 2\,GHz in full
polarization. (c) And finally, we aim to cover as many radio
recombination lines as possible (19) at intermediate spectral
resolution because the ionized lines have broader thermal line width
(on the order of 20\,km\,s$^{-1}$). The spectral set up of the survey,
shown in Table \ref{setup}, was decided based on these goals. The
setup was slightly different for the pilot study, phases 1 and 2
(marked as phase p, 1 and 2, respectively, in the table), as a result
of adjustment and optimization of observation strategy after the pilot
study.

\begin{figure}[htb] 
\includegraphics[width=0.49\textwidth]{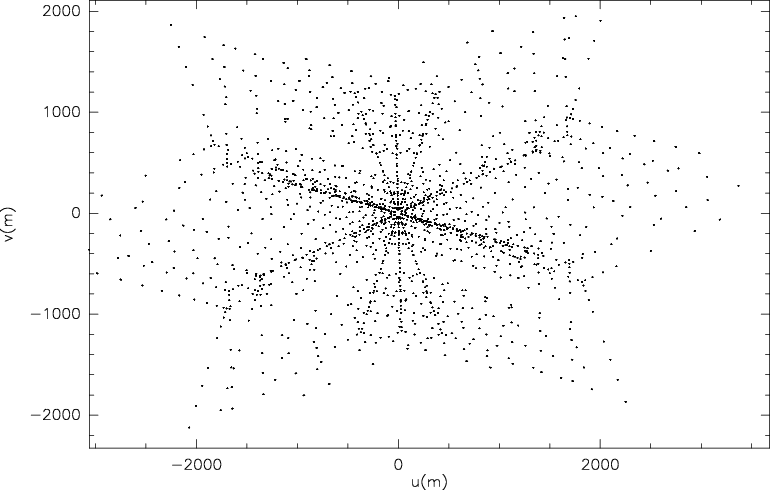}
\caption{Example of uv-coverage of a single THOR pointing within the
  tile around 16.25\,deg longitude in the averaged continuum band
  cont4.}
\label{uv-coverage}
\end{figure}

\begin{table}[htb]
\caption{Spectral setup}
\begin{tabular}{lcccccc}
\hline
\hline
        & cent.\,$\nu$    & width$^a$ & $\Delta \nu$ & $\Delta v$    & phase$^b$ & RFI$^c$ \\
            & (MHz)    & (MHz) & (kHz)       & (km\,s$^{-1}$) &     \\
\hline
HI          & 1420.406 & 2     & 1.95        & 0.41          & p \\
HI          & 1420.406 & 2     & 3.91        & 0.82          & 1,2 \\
OH1         & 1612.231 & 2     & 3.91        & 0.73          & p \\
OH1         & 1612.231 & 2     & 7.81        & 1.46          & 1,2 \\
OH2/3       & 1665.402 & 4     & 3.91        & 0.70          & p \\
OH2         & 1665.402 & 2     & 7.81        & 1.40          & 1$^d$ \\
OH2/3       & 1665.402 & 4     & 7.81        & 1.40          & 2 \\
OH4         & 1720.530 & 2     & 3.91        & 0.68          & p \\
OH4         & 1720.530 & 2     & 7.81        & 1.36          & 1,2 \\
cont1       & 1052     & 128   & 2000        & 570           & p,1,2 \\
cont2       & 1180     & 128   & 2000        & 508           & p,1,2 & x \\
cont3       & 1308     & 128   & 2000        & 459           & p,1,2 \\
cont4       & 1436     & 128   & 2000        & 418           & p,1,2 \\
cont5       & 1564     & 128   & 2000        & 384           & p,1,2 & x\\
cont6       & 1692     & 128   & 2000        & 355           & p,1,2 \\
cont7       & 1820     & 128   & 2000        & 330           & p,1,2 \\
cont8       & 1948     & 128   & 2000        & 308           & p,1,2 \\
H186$\alpha$& 1013.767 & 4     & 31.25       & 9.2           & p \\
H186$\alpha$& 1013.767 & 2     & 15.63       & 4.6           & 1,2 \\
H178$\alpha$& 1156.299 & 4     & 31.25       & 8.1           & p \\
H178$\alpha$& 1156.299 & 2     & 15.63       & 4.1           & 1,2 \\
H176$\alpha$& 1196.028 & 4     & 31.25       & 7.8           & p \\
H176$\alpha$& 1196.028 & 2     & 15.63       & 3.9           & 1,2 \\
H175$\alpha$& 1216.590 & 4     & 31.25       & 7.7           & p \\
H173$\alpha$& 1259.150 & 2     & 15.63       & 3.7           & 1,2 & x\\
H172$\alpha$& 1281.175 & 4     & 31.25       & 7.3           & p & x\\
H172$\alpha$& 1281.175 & 2     & 15.63       & 3.7           & 1,2 & x\\
H171$\alpha$& 1303.718 & 4     & 31.25       & 7.2           & p \\
H171$\alpha$& 1303.718 & 2     & 15.63       & 3.6           & 1,2 \\
H170$\alpha$& 1326.792 & 4     & 31.25       & 7.1           & p \\
H170$\alpha$& 1326.792 & 2     & 15.63       & 3.5           & 1,2 \\
H169$\alpha$& 1350.414 & 4     & 31.25       & 6.9           & p & x\\
H169$\alpha$& 1350.414 & 2     & 15.63       & 3.5           & 1 & x\\
H168$\alpha$& 1374.601 & 4     & 31.25       & 6.8           & p \\
H168$\alpha$& 1374.601 & 2     & 15.63       & 3.4           & 1,2 \\
H167$\alpha$& 1399.368 & 4     & 31.25       & 6.7           & p \\
H167$\alpha$& 1399.368 & 2     & 15.63       & 3.4           & 1,2 & x\\
H166$\alpha$& 1424.734 & 4     & 31.25       & 6.6           & p \\
H166$\alpha$& 1424.734 & 2     & 15.63       & 3.3           & 1,2 \\
H165$\alpha$& 1450.716 & 4     & 31.25       & 6.5           & p \\
H165$\alpha$& 1450.716 & 2     & 15.63       & 3.2           & 1,2 \\
H161$\alpha$& 1561.203 & 2     & 15.63       & 3.0           & 1 & x\\
H158$\alpha$& 1651.541 & 2     & 15.63       & 2.8           & 1,2 \\
H156$\alpha$& 1715.673 & 2     & 15.63       & 2.7           & 1,2 \\
H155$\alpha$& 1748.986 & 2     & 15.63       & 2.7           & 1,2 \\
H154$\alpha$& 1783.168 & 2     & 15.63       & 2.6           & 1,2 \\
H153$\alpha$& 1818.246 & 2     & 15.63       & 2.6           & 1,2 \\
H152$\alpha$& 1854.250 & 2     & 15.63       & 2.5           & 1,2 \\
H151$\alpha$& 1891.212 & 2     & 15.63       & 2.5           & 2 \\
\hline
\hline
\end{tabular}
{\footnotesize
~\\
$^a$ Bandwidth of the individual spectral units.\\
$^b$ The flags correspond to the pilot project (p), and phases 1 and 2.\\
$^c$ These flags mark spectral windows that were unusable because
of RFI.\\
$^d$ In phase 1, we unfortunately missed the 1667 line. For the pilot and phase 2, the 1667 line is included in the OH2/3 setup at 1665\,MHz.}
\label{setup}
\end{table}

\section{Observations, calibration, and imaging} 
\label{obs}

\subsection{Observations}
\label{obs_sub}

Except for the pilot study, which covered approximately an area
of four square
degrees around W43, we split the observations into tiles of $\sim
1.25\times 2.5$\,deg in Galactic longitude and latitude. The
C-configuration covers baselines between $\sim$40 and $\sim$3400\,m,
which results in angular resolution elements of about $20''$ in the
1--2\,GHz band (radio L-band, Table \ref{res_noise}).
% To achieve an angular resolution of $\sim 20''$ in L-band, the VLA
% was used in its C-configuration.
To be more specific, the pilot study mapped approximately $2\times 2$
degrees centered on the W43 star formation complex between
longitudes of 29.2 and 31.5\,degrees. We mosaicked the area with 59
pointings in a hexagonal mosaic geometry sampled at $\sim 17.9'$
spacing, corresponding to half the primary beam size at $\sim
1.26$\,GHz. Each field was observed 4 $\times$ 2\,min to achieve a
relatively uniform uv-coverage (Fig.~\ref{uv-coverage}). Including
overheads for flux, bandpass, and gain calibration, ten hours were
needed for this part of the project, split into two observing blocks of
five hours each.

While the above approach already provided a smooth coverage, we
optimized it for the rest of the survey. Considering that the primary
beam changes from $45'$ at 1\,GHz to $22.5'$ at 2\,GHz, we used a
rectangular mosaic sampling of $15'$ in Galactic longitude and
latitude, respectively, corresponding to half the primary beam at
1.5\,GHz. Each tile of $1.25\times 2.5$ square degrees in Galactic
longitude and latitude was now covered by a regularly spaced mosaic of
45 pointings. Neighboring tiles had exactly the same $15'$ separation
to have a uniform coverage over the full survey. Figure \ref{mosaic}
gives an example of our mosaic pattern for three neighboring tiles
where the mosaic pattern is shown for the primary beam size of $30'$
at 1.5\,GHz. With the varying primary beam size with frequency, the
theoretical sensitivity varies slightly over the fields. However, on
the one hand, this effect is very small (see also \citealt{bihr2016}),
and on the other hand, our noise limit is not the thermal noise, but it
is dominated by the side lobe (see Sect. \ref{additions}), in particular
for the strong emission from the continuum and the masers.  The phase
1 coverage in longitudes from 14.5 to 29.2, from 31.5 to 37.9, and
from 47.1 to 51.2 degrees was covered in 20 tiles observed for five hours
each. The remaining part of the survey fills the longitude gap between
37.9 and 47.1 degrees and extends from 51.2 to 67.4 degrees in
additional 21 tiles. To obtain good uv-coverage, each pointing was covered three times for approximately
two\,minutes within such a five-hour observing block, and $\sim 50$\,min were needed for flux, bandpass, and gain
calibration.

\begin{figure}[htb] 
\includegraphics[width=0.49\textwidth]{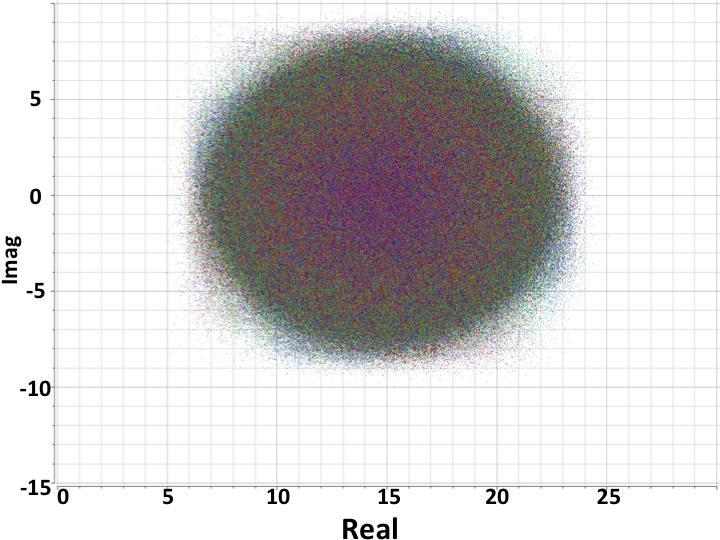}\\
\includegraphics[width=0.49\textwidth]{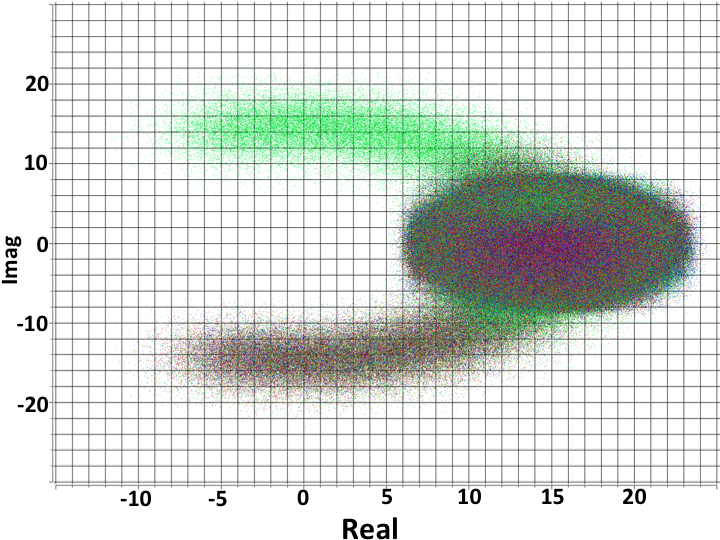}
\caption{Example of diagnostic plots during the calibration process
  (taken from \citealt{bihrphd}). The imaginary part of the
  visibility is plotted as a function of the real part for the
  calibrator 3C286 in the HI line. The color-coding corresponds to the
  different antennas.  For a point source like the calibrators a
  roundish cloud as shown in the top panel is expected. The deviations in
  the bottom panel isolate bad antennas that need to be flagged.}
\label{mickymouse}
\end{figure}

\subsection{Calibration}

The full survey was calibrated with the
CASA\footnote{http://casa.nrao.edu} software package. To
calibrate the pilot study and phase 1 of the survey, CASA
version 4.1.0 and a modified VLA pipeline version 1.2.0 were used. The second half of the survey was calibrated with
slightly newer versions (CASA 4.2.2, pipeline 1.3.1), but
differences between these two versions for the calibration are
minimal. One major problem in L-band data analysis is the undesired
man-made terrestrial signal known as radio frequency interference
(RFI). While some strong RFI and bad antennas were flagged manually
before the calibration, the VLA pipeline also applies automated RFI
flagging on the calibrators during the calibration to improve the data
quality and calibration solutions. Additional RFI flagging was applied later during the image process (see Sect.
\ref{rfi_section}).

Flux, bandpass, and polarization calibration was conducted for all
fields with the quasar 3C286. Two different complex gain calibrators
were used: J1822-0938 for the observing blocks between 14.5 and 39.1
degrees (including the pilot study), and J1925+2106 for the remaining
fields at longitudes $>39.1$ degrees.

After RFI flagging of the calibrators, bandpass, flux, and gain
calibration was applied using standard procedures. The absolute flux
calibration uncertainty at these wavelengths is within $\sim$5\%. No
Hanning smoothing was performed, and the weights were not recalculated (CASA
command statwt) because that sometimes affects particularly bright
sources. Some modifications to the pipeline were implemented by us to
improve the quality checking. The calibration was made iteratively,
where after a full calibration additional quality checks and flags were
applied, after which the calibration was conducted again. Figure
\ref{mickymouse} presents examples of diagnostic plots where the
imaginary part of the visibility is plotted against the real part for
the calibrator 3C286, which corresponds to the phase and amplitude of
the visibility, respectively. For such a point source a roundish
distribution is expected (Fig.~\ref{mickymouse}, top panel). However,
bad baselines are easily identified by strong outliers from this
roundish cloud (Fig.~\ref{mickymouse}, bottom panel). We
typically iterated the pipeline for each tile two to three times.

\subsubsection{Polarization data}
\label{polarization}

The THOR polarization survey constitutes a significant step forward in terms
of spectral and angular resolution over a larger bandwidth than
previous Galactic plane surveys that include polarization
\citep{condon1998,mcclure-griffiths2001,landecker2010}. The survey
region connects that of the Canadian Galactic Plane Survey to that of
the VLA Galactic plane survey. Compared with previous surveys, the VLA
C-configuration filters out much or all of the diffuse Galactic
  polarized emission. THOR is only sensitive to structure on angular
  scales smaller than $\sim 3\arcmin$ (9\,pc at a distance of
  10\,kpc), which is well suited for more distant supernova remnants
  that most likely populate the narrow latitude range of the
  survey. Since Faraday rotation measure synthesis is performed on the
  ratios $Q/I$ and $U/I$, it is less sensitive to missing short
  spacings than for example deriving a spectral index. The main
science applications of the polarization survey therefore focus on
polarized extragalactic sources and Galactic supernova remnants.

\begin{figure*}[htb] 
\includegraphics[width=0.99\textwidth]{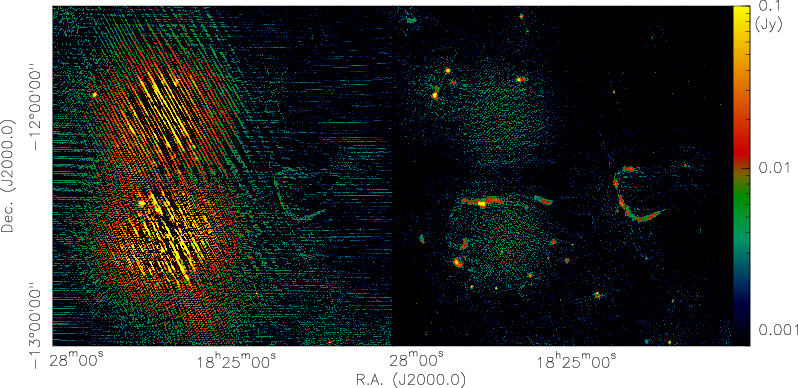} 
\caption{Example of image affected by RFI in the continuum band cont8
  around 1948\,MHz. The left and right panels show data around
  Galactic longitudes of 18\,deg before and after RFI flagging.
    The stripy features in the left panel show the strong RFI
    contribution, which is largely (but not entirely) removed after the
    automatic RFI flagging in the right panel. The color scale is
    chosen to also show the remaining artifacts.}
\label{rfi}
\end{figure*}

Polarization calibration is performed per channel in casa after
bandpass, flux, and gain calibration. Polarization angle calibration is
derived from 3C286. Solutions for instrumental polarization are
derived from the phase calibrator that was observed during
the observing session. Even after calibration for the center of the
field, instrumental polarization increases substantially with distance
from the field center. Off-axis polarization calibration requires a
correction to Stokes Q and U that depends on the location in the field
and to first order is proportional to total intensity. Pending new
holography measurements and implementation of off-axis polarization
calibration in casa, our polarization calibration only applies to the field
center. We intend to process the data including off-axis
polarization calibration before they are released. However, experiments
on bright thermal emission, for instance, on ultracompact H{\sc ii} regions,
indicate that leakage in our mosaics is restricted to cases with
fractional polarization $\lesssim 1\%$ and very little Faraday
rotation \citep[as argued by][]{giessuebel2013}. Because of the
  significantly larger calibration and imaging requirements for the
  polarization data, even the first half of the survey is not fully
  reduced yet. Therefore, polarization data will be provided
  successively (see Sect. 4.4).

The visibility data are imaged in Stokes $I$, $Q$, and $U$ in 2\,MHz
channels across the observed frequency range from 1 to 2 \,GHz. The
restoring beam was calculated for each channel to preserve
the native angular resolution. This allows us the flexibility to
analyze the upper part of the frequency band separately at higher
angular resolution. The lowest-frequency spectral window is noisier by
a factor $\sim 2$ but is important for polarization science
because it nearly doubles the coverage of the survey in $\lambda^2$
space as the adjacent spectral window is lost to RFI.

\subsection{Imaging}

The imaging process for THOR is by far the most computationally
intensive and time-consuming process of the data reduction for several
reasons: (a) Large areal coverage combined with good angular
resolution results in very large images with thousands of pixels in
each spatial axis. The pixel sizes were adapted to the frequencies and
typically had $\sim$4-5 pixels per resolution element in one dimension.
(b) Thirty spectral bands (Table \ref{setup}) produce an enormous
bandwidth at high spectral resolution, that is, many frequency channels,
which needs to be imaged one after the other. (c) The CASA software
package works with high input/output rates of the data to disk and
back, which is a strong constraint on the computing resources. While
the raw data already comprise approximately 4\,TByte,
the total data volume after imaging is increased by up to a factor 10.

The computing process can be optimized by shared file systems such as
the lustre system existing at the NRAO site in Socorro. Based on the
requirements for THOR and future programs, we acquired a comparable
computing cluster with a similar shared file system (FHGSF:
Frauenhofer shared file
system\footnote{http://www.beegfs.com/content/}), which improves the
imaging speed significantly. Nevertheless, a single spectral data cube
covering $\sim 200$\,km\,s$^{-1}$ at a spectral resolution of
1.5\,km\,s$^{-1}$ for one tile ($1.25\times 2.5$\,deg) needs
approximately two weeks for the imaging process. Imaging all tiles and
spectral windows like this in sequence would be prohibitively time
consuming. However, with the new computing system, the imaging of the
data becomes feasible, although it still takes several years for the
full survey, in particular because a large part has to be imaged
several times due to tests and software improvements. The actual
multiscale cleaning, embedded in the CASA task CLEAN, is used for
imaging, and the details are described in Sect. \ref{additions}.

\subsubsection{Different CASA versions}

All data shown are imaged with the CASA version 4.2. Although versions
4.3 to 4.6 became available during the data reduction time, we tested
these versions, and for our large-scale mosaics, 4.2 resulted in the
best and most reliable results. The test reference images were made
with isolated sources close to the center of an individual VLA
pointing without using the mosaic algorithm, but where these
individual pointings were significantly offset from the phase center
of a 45 pointing mosaic tile. Imaging such an individual pointing in
CASA 4.2 and the following versions gave identical results, which is
also expected and should be the case. Therefore, we considered the
measured fluxes of these images as reliable and used them as reference
values. Imaging the corresponding 45 pointing mosaic tile gave
slightly varying fluxes for the same sources within 4.2 and 4.3 to
4.6. The variations within 4.2 were usually below
10\%, whereas in 4.3 to 4.6 we saw systematic flux density deviations
with distance from the phase center of up to 20\% or more in the
mosaics. The main reasons for these discrepancies are rooted in the
implementation of the corrections for the primary beam response and
the distance to the phase centers, which were not properly accounted
for in 4.3 to 4.6 (Kumar Golap, priv.~comm.). Furthermore, the primary
beam correction for the continuum images was calculated in 4.3 and 4.4
for the first channel of a continuum image and for 4.2 for the central
channel. Taking these variations into account, we so far consider CASA
4.2 as the most reliable version for the purpose of our large mosaic
and wide-field imaging. A future data release may have the whole
survey data re-imaged with a newer version of CASA as and when these
problems are sorted out.

\subsubsection{Radio frequency interference (RFI)}
\label{rfi_section}

Radio frequency interference is a severe problem at the long
wavelengths of this survey. While a few frequency ranges, in
particular around the HI and OH lines, are protected and do not have
strong RFI problems, many of the other spectral windows are strongly
affected by it. In particular, two of the continuum spectral windows
and seven of the recombination line spectral windows are so strongly
affected by RFI that we excluded them from the analysis entirely. These
windows are marked in Table \ref{setup}.

Automated RFI flagging algorithms are very useful, but they also need
to be used with caution. In particular for strong narrow spectral
lines, for example, the OH masers, RFI flagging algorithms are prone to also
flag the maser peaks. Hence, for the HI and OH spectral windows we
refrained from directly applying automated RFI flagging
algorithms. Since these windows are protected anyway, however,
it was not a
severe problem, and individual checks and some sparse flagging
by hand accounted for most of the RFI in these bands.

For the continuum bands, the spectral windows cont4 and cont7
around 1.4 and 1.7\,GHz are of the best quality. The spectral window
cont4 around the HI line has hardly any RFI and requires little or no
RFI flagging. The window cont7 around 1.8\,GHz still is of good
quality, and only a few individual RFI checks and flagging accounted
for most of the RFI features there. All other continuum spectral
windows were more strongly affected by RFI. Since this RFI
contamination varies in frequency, sky position, and time,
manual flagging is not feasible for this large survey. We
therefore explored the RFlag algorithm within CASA that was first introduced to
AIPS by E.~Greisen in 2011. This iterative algorithm considers the
statistics in both time and spectral domain. Outliers are identified
by considering each individual spectral channel for the whole duration
of an observation of a target, and also by considering all spectral
channels for each integration time step, and the flagging of outliers
is made iteratively (see CASA
manual\footnote{http://casa.nrao.edu/docs/cookbook/index.html} for
more details). As mentioned above, strong spectral features such
as masers would be considered outliers and thus flagged during that
procedure. Therefore, the RFlag algorithm cannot be applied to the HI
and OH spectral windows. However, since the continuum emission should
not show such variations over the bandpass, it can be applied to
those windows. Furthermore, the radio recombination lines are
typically extremely weak and can often only be reliably identified
after stacking several lines (see description below). Hence, the RFlag
algorithm can also be applied on weak features such as the
RRLs. Testing the performance of RFlag showed that using the default
thresholds ($5\sigma$ rms in time and frequency domain) reliably
removes most of the RFI in the respective windows (most, but not
  all, see Fig.~\ref{rfi} for possible residuals). Bands like cont2 around
1.2\,GHz have so much RFI over the whole bandpass that no usable data
remain, hence the algorithm does not give reasonable results in such
extreme cases. However, for the non-marked spectral windows shown in
Table \ref{setup}, the RFlag results greatly improved the data
quality. Figure \ref{rfi} shows one example region where the RFI was
largely removed by applying the RFlag algorithm once with these
default values.

We performed several tests with the RFlag algorithm to further quantify its
effects. We imaged a dataset in the RFI-free cont4 band around the HI
line with and without applying the RFlag algorithm. Measuring the
noise in both images, it is almost identical, indicating that
the RFlag algorithm did not flag significant good data. Extracting
the flux densities using these two approaches, we find no significant
deviations over the full range of flux densities for unresolved and
small sources (smaller than $100''$). For more details and
corresponding figures, see the continuum catalog paper by
\citet{bihr2016}. The RFI removal effects for larger-scale structures
are discussed in the following subsection.

\subsubsection{Spatial filtering for extended sources}
\label{spatial_filter}

Extended sources suffer from spatial filtering effects for
several reasons. While the observations clearly filter out emission
because of the missing short baselines, RFI removal also lowers
flux density measurements of extended sources. 

Within a full 12-hour track, the VLA can observe spatial scales up to
970$''$ in L-band and C-configuration. However, with the shorter integrations
of $\sim$6\,min per mosaic pointing, flux density on smaller scales is
also filtered out. To investigate the spatial filtering in our setup,
we simulated observations with Gaussian intensity profiles of varying
sizes, employing our given uv-coverage. We find that sources with sizes
up to $120''$ are reasonably well recovered, with flux density losses
of less than 20\%. The use of the multiscale algorithm (see also next
section) within CASA was very important for recovering the flux
density. Since the filtering also depends on frequency, spectral
indices are not reliable for structures
larger than $120''$ . Nevertheless,
the spectral indices for smaller structures remain trustworthy (see
also \citealt{bihr2016}).

The additional spatial filtering due to the RFI flagging is also based
on the fact that extended sources show high amplitudes and amplitude
gradients at small uv-distances. Since the RFlag algorithm flags
outliers in the time and frequency domain, some of these high values
may also be considered as outliers and hence be flagged. Similar to the
RFI tests outlined in the previous selection, we again applied the
RFlag algorithm to the cont4 spectral window around the HI line that
is practically RFI free. As shown in \citet{bihr2016}, for
uv-distances smaller than 300$\lambda$, this flagging starts to
significantly affect the flux densities, and more than $70\%$ of the data may
be flagged. Following the simple estimate of the corresponding spatial
scale $\theta=\lambda/D$ with $\theta$ the angular scale and $D$ the
baseline length, $300\lambda$ corresponds to scales of $\sim
600''$. For structures on scales between $100''$ to $300''$, the flux
density removal due to RFlag is only on the order of $5-10\%$. Hence,
while for small spatial scales the flux density uncertainties because
of spatial filtering are negligible, they have to be considered
more seriously for the
large-scale structures.

In summary, while automated RFI removal does affect the flux density
on larger scales, the spatial filtering is dominated by the normal
filtering of the interferometer arising from the missing short
spacings.

\begin{figure*}[htb] 
\includegraphics[width=0.99\textwidth]{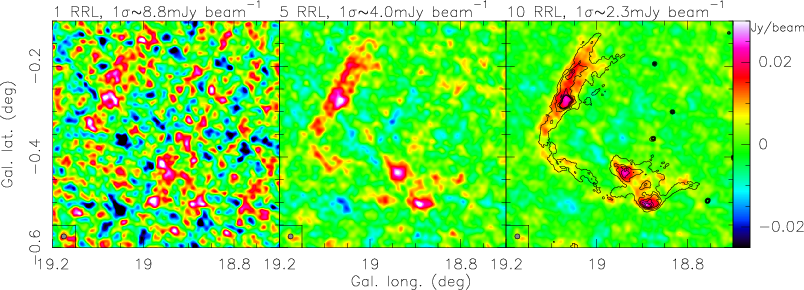}
\caption{Example of recombination line stacking toward the H{\sc
      ii} region G18.9-0.3. The images show one recombination line
  emission region for various numbers of stacked images in a single
  channel of 10\,km\,s$^{-1}$ width (around 70\,km\,s$^{-1}$). The
  left panel represents a single recombination line (H$152\alpha$),
  the middle contains 5 RRLs, and the right panel combines 10
  RRLs. The synthesized beam of $40''$ is shown in the bottom
left corner of
  each panel. For reference, the contours in the right panel present
  the corresponding 1.4\,GHz continuum data (see also
  Fig.~\ref{continuum2}) with contour levels from 50 to 150\,K in steps
  of 20\,K.}
\label{stacks}
\end{figure*}

\subsubsection{Additional imaging effects and procedures}
\label{additions}

In addition to the imaging procedures and effects outlined above, a
few more important steps are also described below.

\paragraph{Multiscale cleaning:} The clean algorithm in its standard
form identifies peak emission features and hence better
recovers smaller point-like emission. Although it is possible to
recover some larger-scale emission by adding many point-like
components, it nevertheless is not ideally suited for extended
emission. The CASA package provides a multiscale version of this
procedure \citep{rau2011}. As the name suggests, it does not look only
for point-like components, but also considers larger spatial
scales. The CASA setup allows us to select different spatial scales to
focus on, and we experimented with many of them using mock input
images. After extended testing, the default values for the multiscale
clean task were used. In this setup, the algorithm focuses on
recovering three main spatial scales: point-sources, the synthesized
beam size, and three times the synthesized beam size. This setup
enables us to recover the large-scale emission best. Therefore, we
used this setup during the imaging process. We cleaned the data using
a robust value in the clean-task of CASA of 0.5. The positional
uncertainties for the final images are within $2''$ \citep{bihr2016}.

\paragraph{Combination with VGPS data:} The THOR survey was initially
set up to be directly combined with the HI and 1.4\,GHz continuum data
from the VGPS survey \citep{stil2006}. This survey observed the HI
line with a spectral coverage of effectively $\sim 1.62$\,MHz (or
341\,km\,s$^{-1}$), and the line-free part of this spectrum was used
for the continuum image. VGPS combined VLA D-configuration configuration data
with single-dish observations from the GBT and Effelsberg. The angular
resolution of their final data product is $60''$, and because of the
combination of VLA with GBT, the large-scale emission is
recovered. These data are used by THOR to complement the short spacing
information for the HI and the 1.4\,GHz continuum
emission. While this works very well for the HI data because both data
are on the same spectral grid, we recall that for the
continuum totally different widths of the spectral bands are used
($\sim 120$\,MHz versus $\sim 1$\,MHz). Therefore, while the spatial
structure of the combined 1.4\,GHz continuum image is reliable, the
absolute fluxes need be considered with caution. We therefore
used the THOR-only data
to determine our
spectral index \citep{bihr2016}.

The THOR and VGPS data can be combined in two different
ways. Either the VGPS data are used as an input model in the
deconvolution of the THOR data, or the THOR data
are independently imaged andare  afterwards combined with the feather task in
CASA. We compared these two approaches, and while using VGPS as an
input model gave good results for the 1.4\,GHz continuum data, this
approach reduced the signal-to-noise ratio for the HI data. Therefore,
for the latter we first imaged the THOR HI data separately and
then feathered them with the VGPS data.

\paragraph{Recombination line imaging and stacking:} The treatment of
the radio recombination lines (RRLs) of this survey is unique because
each individual line is too weak to be detected in most
regions. However, because of the power of the new VLA correlator, we
were able to observe 19 of them simultaneously. While some of the spectral
windows are RFI contaminated (see Table \ref{setup}), typically about
12 spectral windows were usable. Because of the weak emission in each
individual RRL, cleaning the data was not appropriate. For the
RRLs we therefore Fourier transformed the data and afterward worked with the
dirty images. All spectral lines were imaged with exactly the same
velocity resolution of 10\,km\,s$^{-1}$. After smoothing all RRL
images to the same angular resolution ($40''$, corresponds
approximately to the poorest resolution achieved at the longest
wavelength), we then stacked the images with equal weights in the
velocity domain to improve the signal-to-noise ratio of the final RRL
images.  After this stacking process, the averaged RRL emission could
be recovered in a significant number of regions in the survey (Sect.
\ref{rrl}). Figure~\ref{stacks} presents in an example toward the H{\sc ii}
region G18.9-0.3 how the stacking process improves the noise and thereby
the signal-to-noise ratio. Measuring the rms in a single
10\,km\,s$^{-1}$ channel results in values of 8.8, 4.0, and
2.3\,mJy\,beam$^{-1}$ for the 1, 5, and 10 stacks of this example
region, respectively.

Although the native spectral resolution for most parts of the survey
is better than 5\,km\,s$^{-1}$, to image such a large number of
datasets in a uniform manner and to achieve an unbiased census of RRL
detections, we are currently using the data at a spectral resolution
of 10\,km\,s$^{-1}$. However, considering that the thermal line-width
of ionized gas at 10000\,K is $\sim$15\,km\,s$^{-1}$ and typically
measured line-widths of recombinations lines in H{\sc ii} regions
between 20 and 25\,km\,s$^{-1}$ \citep{anderson2011}, this is still
reasonable.

\paragraph{Angular resolution and rms noise:} The final angular
resolution elements and the rms noise levels vary between the
different lines and continuum bands, but also from the lower to the
upper end of the bandpass and from the low- to the
high-longitude range. Table \ref{res_noise} summarizes the main
angular resolution and mean noise parameters of THOR. While in
emission-free regions we almost reach the thermal noise, our maps are
mostly dominated by side-lobe noise. This means that the noise levels vary
throughout the fields. This behavior is particularly strong in the
continuum emission and around the strong OH maser. The
side-lobe noise is a direct reflection of the uv-coverage and the
corresponding dirty beam. For a very uniform uv-coverage, the dirty
beam is almost Gaussian and side-lobe noise is weak. However, for less
uniform uv-coverage, the dirty beam has stronger negative features
that remain difficult to clean. As described in Sect. \ref{obs_sub}
and shown in Fig. \ref{uv-coverage}, to conduct such a large survey,
the uv-coverage of each individual pointing does not fill the uv-plane
well. This directly results in a less perfect dirty beam and thereby
higher side-lobe noise as observed here. This effect also depends on
the strength of the sources and is therefore particularly prominent
for the strong masers and the strongest continuum sources. For
the other parts of the survey, side-lobe noise is less severe and can
almost be neglected.

  \citet{bihr2016} analyzed the noise behavior of the continuum data
  in the first half of the survey in depth, and found that 50\%
  of the survey area have a noise level below a 7\,$\sigma$ level of
  3\,mJy\,beam$^{-1}$. Above their chosen 7\,$\sigma$ threshold 95\%
  of all artificially injected sources are detectable. Only 10\% of
  the area has a noise level above a 7\,$\sigma$ value of
  8\,mJy\,beam$^{-1}$. For comparison, \citet{walsh2016} analyzed the
  completeness of the OH maser data in the pilot field around W43, and
found that almost all OH data are complete at
  $\sim$0.25\,Jy\,beam$^{-1}$, and that 50\% are complete at a level
  of 0.17\,Jy\,beam$^{-1}$.

  The side-lobe noise is less of a problem for the HI because these
  data are complemented by VGPS D-configuration and GBT data, which
  significantly improves the image quality. Nevertheless, side-lobe noise has also to be taken
into
  account for the HI data
  toward strong continuum sources. Since the
  recombination lines are very weak, they also do not suffer much from
  side-lobe noise compared to these lines
and the continuum. Furthermore, as outlined in the previous paragraph,
  the signal-to-noise ratio increases because of the stacking. In
  total, the rms for the recombination lines is much more uniform and
  represented well by the value shown in Table \ref{res_noise}.

In the appendix (Sect. \ref{noise_maps}) we show representative
noise maps for the continuum, HI, and OH for a selected tile of the
survey to visualize the noise spread.

\begin{table*}[htb]
\caption{Angular resolution, rms noise (both for THOR-only data), and
  data products$^a$}
\label{res_noise}
\begin{tabular}{lrrr}
\hline
\hline
Band & $\theta^b$ & rms & data products$^b$ \\
     & $('')$  & $\left(\frac{\rm{mJy}}{\rm{beam}}\right)$ \\
\hline
HI & $15.9''\times 12.8''$ to $19.9''\times14.2''$ & 12$^c$ & THOR only with \& without cont., THOR merged with VGPS continuum-subtracted \\
OH1 & $13.5''\times 13.2''$ to $18.7''\times12.5''$ & 10$^d$ & continuum-subtracted, native resolution and $20''$ smoothed\\
OH2/3 & $13.1''\times 12.9''$ to $18.1''\times12.1''$ & 10$^d$ & continuum-subtracted, native resolution and $20''$ smoothed \\
OH4 & $12.7''\times 12.4''$ to $17.6''\times11.8''$ & 10$^d$ & continuum-subtracted, native resolution and $20''$ smoothed\\
cont1 & $16.5''\times 15.7''$ to $24.4''\times 15.1''$ & 1.1$^e$ & native resolution and $25''$ smoothed\\
cont3 & $13.1''\times 12.3''$ to $19.7''\times 12.5''$ & 0.7$^e$ & native resolution and $25''$ smoothed\\
cont4 & $12.6''\times 11.9''$ to $18.1''\times 11.1''$ & 0.6$^e$ & native resolution and $25''$ smoothed\\
cont6 & $10.5''\times 9.9''$ to $15.4''\times 9.1''$ & 0.6$^e$ & native resolution and $25''$ smoothed\\
cont7 & $10.0''\times 9.7''$ to $14.5''\times 8.9''$ & 0.5$^e$ & native resolution and $25''$ smoothed \\
cont8 & $9.0''\times 8.3''$ to $13.1''\times 8.1''$ & 0.7$^e$ & native resolution and $25''$ smoothed\\
H$n\alpha$ & $40''$ & 3$^f$ & stacked images with 10\,km\,s$^{-1}$ and $40''$ resolution\\
 & & & for all products also calibrated visibilities\\
\hline
\hline
\end{tabular}
{\footnotesize ~\\
  $^a$ This data release 1 contains the data of the first half of the survey. The remaining data will follow after the ongoing calibration and imaging process.\\  
  $^b$ The synthesized beams depend on Galactic longitude.\\  
  $^c$ rms per channel after smoothing to a uniform beam of $21''\times 21''$.\\
  $^d$ rms per channel after smoothing to a uniform beam of $20''\times 20''$.\\
  $^e$ This is the rms in a emission-free region. Considering side-lobe noise, 90\% of the whole coverage is below $\sim$1.6\,mJy\,beam$^{-1}$ \citep{bihr2016}.\\
  $^f$ After smoothing to $40''$ in 10\,km\,s$^{-1}$ channels of stacked maps.
}
\end{table*}

\subsection{Data products and data access}
\label{products}

The data are provided to the community as calibrated images. While the
continuum data for each band (Table \ref{res_noise}) are accessible as
individual single-plane fits-files, the spectral line data are
provided as fits data-cubes. These data cubes always cover the whole
velocity range of Galactic emission in the respective part of the
Milky Way. To stay within reasonable file-size limits, the data can be
downloaded in tiles of approximately 2.5 square degrees each. We
provide the data with different angular resolution and with and
without continuum-subtraction. More details can be found in Table
\ref{res_noise}. Currently, only full tiles can be downloaded, but we plan also smaller cutout-image options. The
current data release 1 contains the data of the first half of the
survey, the second half will follow after the ongoing data processing
has been completed. Because of the significantly larger data
calibration and imaging requirements for the polarization, which is not
completed yet (Sect. \ref{polarization}), the polarization data will
also be provided at a later stage. The data can be accessed at the
project web-page http://www.mpia.de/thor.

\section{Initial results}

\subsection{Continuum emission}

THOR provides a variety of continuum data products. We have the full
spectral coverage from 1 to 2\,GHz, which enables us to derive spectral
indices for all identified regions. These spectral indices are a very
useful tool for differentiating the physical properties of the regions,
for instance, for resolving extragalactic synchrotron emission from Galactic H{\sc
  ii} regions. For the first half (phase 1) part of THOR, a detailed
presentation of the continuum source catalog and early results can
by found in \citet{bihr2016}, and we refer to that paper for more
details.

\begin{figure}[htb]
\begin{center}
  \includegraphics[width=0.48\textwidth]{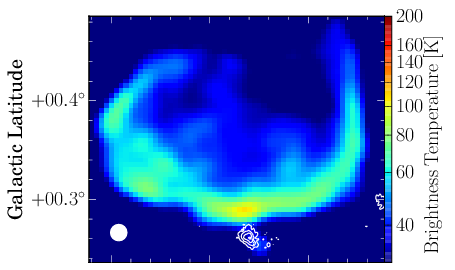}\\
  \includegraphics[width=0.48\textwidth]{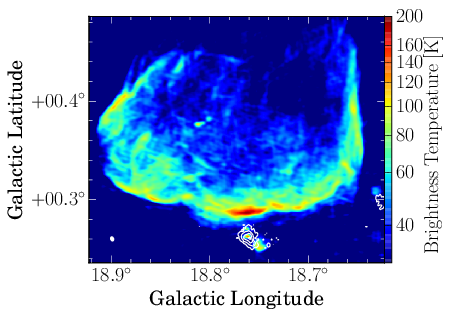}
\end{center}
\caption{\footnotesize The two panels present the 1.4\,GHz
  continuum emission toward the supernova remnant G18.8+0.3. While the
  top panel shows the old VGPS data, the bottom panel presents the
  vast improvement obtained with our new THOR+VGPS data. The contours
  in both panels present the 870\,$\mu$m continuum data from the
  ATLASGAL survey \citep{schuller2009}.}
\label{cont} 
\end{figure} 

In contrast to the other continuum bands that are only observed with
THOR in the C-configuration, for spectral window cont4 we also
have the
complementary VGPS data from the VLA in D-configuration and the
Effelsberg observations.  Although the absolute flux density should be
considered with caution in the combined dataset (Sect.
\ref{additions}), this THOR+VGPS 1.4\,GHz continuum dataset gives a
unique view of our Milky Way. It resolves the small-scale structure
with the THOR data and at the same time recovers large-scale structure
from the VGPS survey. Figure \ref{cont} shows a zoom into one region (the
supernova remnant G18.8+0.3) where the direct comparison between the
previous VGPS data at $60''$ resolution and our new THOR data at
$15''\times 11''$ resolution is presented. The improvement in angular
resolution and dynamic range is striking. For example, while the
low-resolution VGPS image does not reveal a cm counterpart to the cold
870\,$\mu$m dust emission at the southern tip of the supernova
remnant, the new THOR image clearly reveals an embedded H{\sc ii}
region within the ATLASGAL dust core. Since this source is at the tip
of the supernova remnant, it may suggest triggered star
formation. This is only an example for the
direction in which the new high-quality data can lead the research.

\begin{figure*}[htb] 
\includegraphics[width=0.71\textwidth]{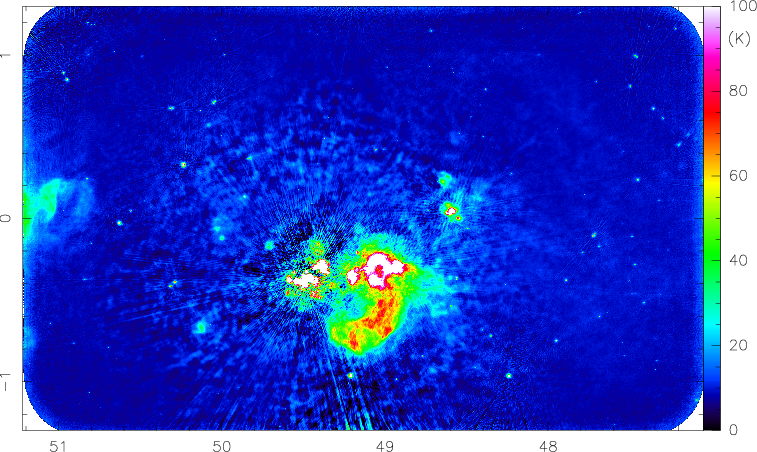}\\
\includegraphics[width=0.99\textwidth]{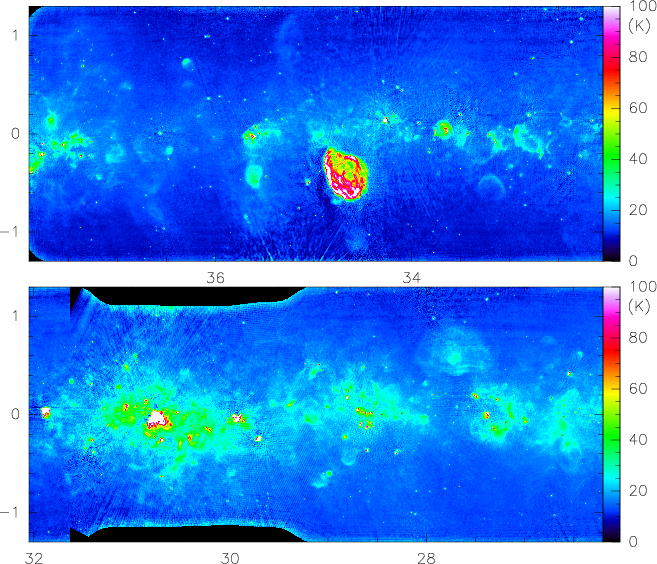}
\caption{THOR continuum data at 1.4\,GHz in degrees of Galactic
  longitude and latitude. The images are constructed by combining the
  THOR C-configuration data with the VGPS data that were produced from
  the VLA D-configuration with the GBT. The angular resolution of this
  image is $20''$. The conversion from Jy\,beam$^{-1}$ to K is made in
  the Rayleigh-Jeans limit. The top panel shows the region around W51
  at the Sagittarius tangent point, and the bottom two panels present
  the areas between longitudes 37.9 and 26.2 degrees. The color
scale
  is chosen to simultaneously show as much large- and small-scale
  emission as possible. The slightly different sky coverage in
  latitude around 31\,degrees longitude is taken from the pilot study.}
\label{continuum1}
\end{figure*}

\begin{figure*}[ht] 
\includegraphics[width=0.99\textwidth]{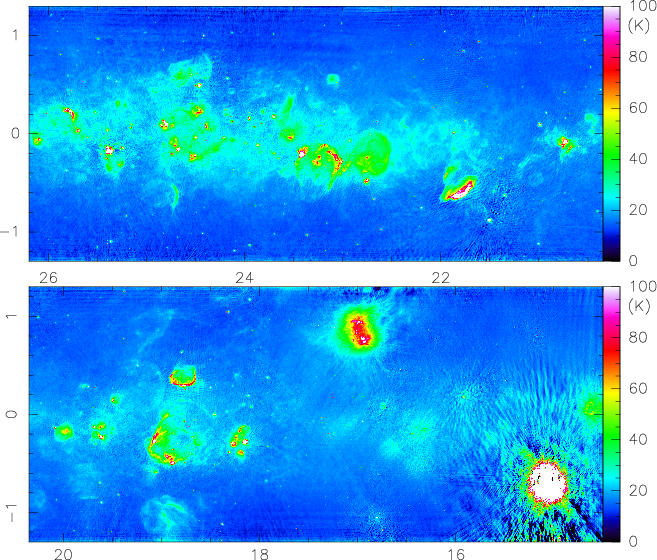}
\caption{THOR continuum data at 1.4\,GHz in degrees of Galactic longitude and
  latitude (continued). The longitudes extend from 26.2 to
  14.5\,degrees. For longitudes below 17.5\,deg, no complementary
  D-configuration data exist (only the single-dish GBT data are available).}
\label{continuum2}
\end{figure*}

Figures \ref{continuum1} and \ref{continuum2}
present these combined images. Below 17.5\,deg longitude, the VLA
D-configuration data do not exist, and we were only able to combine the new THOR data
with the complementary GBT single-dish observations. This resulted in
less structural information in that region and in more side-lobe
noise, in particular around the H{\sc ii} region M17 at $\sim$15\,deg
longitude.

The 1.4\,GHz continuum images exhibit a multitude of features. Close
to the Galactic mid-plane, the emission is dominated by Galactic and
often extended structures. Most of these are either H{\sc ii} regions
or supernova remnants. Comparing these structures with Galactic H{\sc
  ii} regions identified in the mid-infrared by
\citet{anderson2014,anderson2015}, we mostly find good
matches between the radio and mid-infrared identified regions.  The
combined VGPS and THOR 1.4\,GHz data recovers emission from both
extended and compact sources.  Close to the Galactic mid-plane, most
of the emission is from Galactic H{\sc ii} regions (cataloged by
\citealt{anderson2015}), but there is also emission from known
supernova remnants (SNRs).  From visual inspection, we find that
nearly all known H{\sc ii} regions and SNR are detected in these data.

\clearpage 

Furthermore, the THOR+VGPS data have revealed a new H{\sc ii} region and
SNR candidates.  Using WISE mid-infrared data, \citet{anderson2015}
identified over 700 H{\sc ii} region candidates that lacked radio
continuum emission in previous surveys over the area now covered by
THOR data release 1 (Fig. \ref{continuum1}).  H{\sc ii} regions
should all have coincident radio continuum and mid-infrared emission
(e.g., \citealt{haslam1987}).  By visual inspection of the THOR data,
we found that 76 of these previous radio-quiet candidates do indeed
have faint radio continuum emission in the THOR+VGPS data, and a
further 52 have emission in at least one of the individual THOR
continuum sub-bands.  These are therefore probably genuine H{\sc
  ii} regions.  The radio continuum sensitivity of THOR+VGPS is
sufficient to detect H{\sc ii} region emission from single B1 stars to
a distance of 20\,kpc, and these data can therefore help to complete
the census of Galactic H{\sc ii} regions over the survey area.  In
addition to the new H{\sc ii} region detections, we have identified
over 50 SNR candidates in the THOR+VGPS data, and are working to
further characterize the nature of these sources. The spectral index
helps to distinguish between H{\sc ii} regions, which reveal a flat or
positive spectral index and SNRs, which typically reveal a spectral
index around -0.5 (e.g., \citealt{green2014,dubner2015}, but see also
\citealt{bhatnagar2011} for other spectral indices in
SNRs). \citet{bihr2016} used the spectral index information to confirm
four SNR candidates proposed by \citet{helfand2006}, which exhibit
these typical spectral indices.

In addition to these extended sources, the continuum images show
many point sources. These are easy to identify at higher Galactic
latitudes because they are less easily confused with Galactic sources, but we also
find many point sources close to the Galactic plane. Most of these
point sources are of extragalactic origin and can often be identified
by a negative spectral index. \citet{bihr2016} conducted a detailed
analysis of the continuum emission in the first half of the THOR
survey, and they identified $\sim$4400 sources of which $\sim$1200 are
spatially resolved. For $\sim$1800 sources they were able to derive
spectral indices with a distribution peaking at values around -1 and
0. These correspond to steep declining sources mostly of extragalactic
origin, whereas the flat-spectrum sources are largely H{\sc ii}
regions. 

\subsection{Atomic hydrogen}
\label{hi}

The atomic hydrogen data can be used in different ways. On the one
hand, we have the THOR-only C-configuration HI observations, which do not
recover the large-scale emission. They are, however, ideally suited to
measure the absorption profiles at high angular resolution against
Galactic and extragalactic background sources. With these absorption
spectra, we can derive the HI optical depth and from that the HI
column density with high accuracy for several hundred lines of
sight
in our Milky Way. Examples for these absorption spectra can be found
in the pilot study paper about the W43 complex \citep{bihr2015b}.
These HI absorption spectra can also be set into context with the
corresponding OH absorption spectra discussed in Sect.
\ref{oh}. While these optical depth measurements are important for
individual regions, it will also be interesting to interpolate between
these individual data points to create an optical depth map of the
Milky Way. Furthermore, the HI absorption data are important to
distinguish the near- and far-distance ambiguities for kinematic
distances within the Milky Way (e.g., \citealt{ellsworth2015}).

\begin{figure}[htb] 
\includegraphics[width=0.49\textwidth]{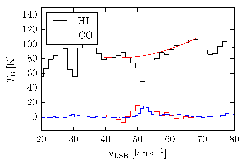}
\caption{HI self-absorption spectrum at the position of l=36.48deg,
  b=-0.04. The spectrum is extracted from the THOR+VGPS data at $40''$
  resolution. The black spectrum shows the original data with a red
  second-order polynom fit to the environmental gas. The red spectrum
  is then the resulting HISA feature used to determine the column
  density (Bihr et al.~in prep.). The blue spectrum is the
  corresponding $^{13}$CO(1--0) emission from the GRS survey
  \citep{jackson2006}.}
\label{hisa}
\end{figure}

The combined THOR+VGPS HI data allow us to also recover the
large-scale emission. However, we recall that the
surface brightness sensitivity in Kelvin gets worse with increasing angular
resolution. For the combined THOR+VGPS data, the $1\sigma$
brightness sensitivity for a spectral resolution of 1.6\,km\,s$^{-1}$
at $21''$, $40''$ , and $60''$ is 16, 3.9, and 1.8\,K, respectively. At
$60''$ resolution, the corresponding $1\sigma$ rms of the VGPS alone
is even slightly superior at $\sim$1.5\,K. When only the large-scale structure is of interest, the VGPS HI
data may still be used, but as soon as higher angular resolution is needed, the
power of the THOR survey can be exploited.

The combined THOR+VGPS data will be also useful in probing intensity
fluctuation of the spectrally resolved HI signal from angular power
spectra over the angular scale range of $\geq20''$ for different parts
of the Galactic plane (e.g.,
\citealt{liszt1993,elmegreen2004,roy2010}). For some of the directions
where the distance-velocity mapping is uniquely known from the
Galactic dynamics, the data cube can also be used to derive the
three-dimensional power spectrum to quantify the structures of atomic
ISM.

One particularly interesting aspect of the HI emission is the
identification and study of cold HI often seen as HI self
absorption or HISA (e.g., \citealt{gibson2005,gibson2005b}). When
features are more  narrow than those of molecular lines (e.g., OH,
C$^{18}$O, $^{13}$CO), this absorption is referred to as HI narrow self-absorption (HINSA, \citealt{li2003}). While the total HI emission is
always a mixture of the cold and warm neutral medium, these HISA
features are dominated by the cold HI component of the ISM. This cold HI is therefore thought to be closely related to the molecular gas during the
formation of molecular clouds. One of the goals of THOR accordingly
is
to systematically study the HISA properties and set them into context
with the even denser portions of molecular clouds visible in CO or
dust continuum emission. HISA features enables us to not only study
the cold HI column density, but also to investigate the kinematic
properties of the HI in comparison to the molecular gas measured in
CO. Figure \ref{hisa} shows an example of a HISA spectrum in
comparison with the molecular component. A detailed study of one
large-scale HISA is also presented in Bihr et al.~(in prep.).

\begin{figure}[htb] 
\includegraphics[width=0.41\textwidth]{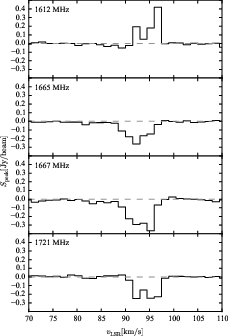}
\caption{Absorption spectra in the OH ground-state transitions
  against a continuum background source ($l = 30.720^o,\; b
  =-0.083^o$). It is unresolved but Galactic, because it is classified as
  a UCH{\sc ii} region in the CORNISH catalog G030.7197-00.0829.  The
  spectra are extracted at the continuum peak position with a velocity
  resolution of 1.5\,km\,s$^{-1}$.}
\label{oh_abs}
\end{figure}

\subsection{OH maser and thermal components}
\label{oh}

The OH part of THOR also covers two very different aspects.  This is
the first unbiased northern hemisphere survey of OH masers in all four
OH transitions at 1612, 1665, 1667, and 1720\,MHz. While some surveys
at lower sensitivity and mainly focusing on the 1612\,MHz maser exist
(e.g., \citealt{sevenster2001}), here we can for the first time
compare the maser properties of the different lines in a statistical
sense. OH masers are known to trace different astrophysical entities. The 1612 OH maser is often associated with evolved stars, for
example (e.g.,
\citealt{sevenster2001}), the 1665 and 1667 masers tend to be more
prominent toward star formation regions (e.g.,
\citealt{reid1981,elitzur1992}), and the 1720 maser is also found
toward supernova remnants (e.g.,
\citealt{elitzur1976,wardle2002}). However, none of these associations
is exclusive, and it is also possible to find all four transitions
toward the same target region (e.g., \citealt{caswell2013,walsh2016}).

A detailed description and statistical analysis of the OH maser
properties toward the four square-degrees pilot region around W43 has been presented by \citet{walsh2016}. The
identification of 103 maser sites in that area covering all four maser
species outlines the great potential of the full survey. In this pilot
region, we identified 72 sites of 1612\,MHz maser emissions, 64\%
of which are associated with evolved stars, 13\% associated with star
formation, and 24\% are of unidentified origin. The 11 maser sites that
emit in the two main lines at 1665 and 1667\,MHz are all located within
star-forming regions. Of the 11 sites with only 1665\,MHz maser
emission, 8 are associated with star formation and three are
of unknown
origin. It is interesting to note that out of the four 1720\,MHz
masers, which are commonly believed to arise in supernova remnants,
three in our field are associated with star formation and the
fourth is again of unknown origin. For more details we refer to
\citet{walsh2016}. The analysis of the remaining part of the survey is
currently being carried out (Walsh et al.~in prep.)

\begin{figure*}[htb] 
\includegraphics[width=0.99\textwidth]{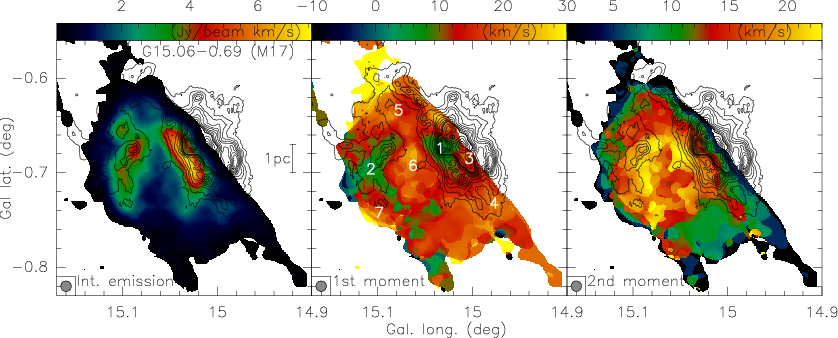}
\caption{Stacked RRL images of G15.1 (also
  known as M17). The left, middle, and right panels present the zeroth-,
  first-, and second-moment maps (integrated emission, intensity-weighted
  peak velocity, and intensity-weighted line width) for the velocity
  range [-40,50]\,km\,s$^{-1}$, respectively. The beam size of
  $40''$ is shown in the bottom left corner of each panel. The contours
  show the 870\,$\mu$m continuum data from the ATLASGAL survey
  starting at a $4\sigma$ level of 200\,mJy\,beam$^{-1}$. The left
  panel gives a linear scale-bar, and the middle panel includes
  numbers indicating the positions of the extracted spectra presented in
  Fig.~\ref{rrl_spectra}.}
\label{rrl_m17}
\end{figure*}

In addition to the maser emission, THOR is also sensitive to the OH
absorption toward strong continuum sources such as H{\sc ii}
regions. These absorption lines are usually of thermal origin and
allow us to study the molecular components of the ISM along the same
lines of sight as the HI absorption lines discussed in Sect.
\ref{hi}. Figure \ref{oh_abs} presents an example of OH absorption
lines toward a bright background continuum source, classified as an
ultracompact H{\sc ii} region in the CORNISH survey catalog
\citep{hoare2012,purcell2013}. The two main lines (at 1665 and
1667\,MHz) and the 1720\,MHz satellite line are found in absorption,
while the OH 1612\,MHz satellite transition is seen in emission. Such
inversion can occur because of radiative maser processes (e.g.,
\citealt{elitzur1992}). The full analysis of the OH absorption lines,
setting them into context with HI absorption spectra, HISA features, and CO emission lines, will be presented in Rugel et al.~(in
prep.).

While most of the absorption spectra are spatially unresolved, we can also
spatially resolve the absorption lines against strong and extended
H{\sc ii} regions  toward
a few particularly strong regions, for example, W43, W51 or M17 (e.g., \citealt{walsh2016}, Rugel et al.~in
prep.). However, for most other parts of the THOR survey, the OH
absorption is mostly very compact. We do not have the
corresponding more D-configuration data (as for the HI or 1.4\,GHz continuum
emission), which are sensitive to the more extended structures, therefore we detect hardly any thermal OH emission. However, we are exploring
whether combining these data with single-dish observations (from
Effelsberg and/or Parkes) will give useful information for studying
the emission of OH on larger scales.

\begin{figure}[htb] 
\includegraphics[width=0.45\textwidth]{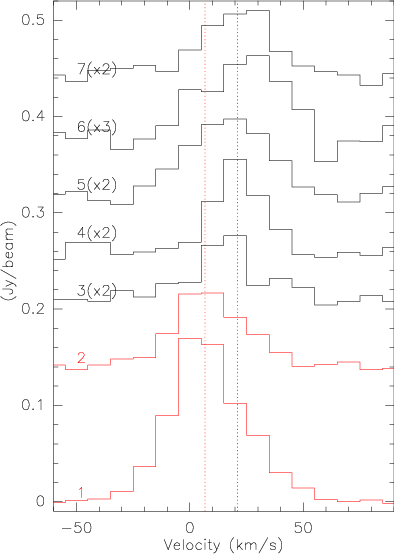}
\caption{Stacked RRL spectra toward the seven
  positions marked in Fig.~\ref{rrl_m17}. The identifying number is
  followed (in parentheses) by the factor by which these have been
  multiplied. The red lowest two spectra are extracted toward two
  intensity maxima, while the remaining black spectra are extracted
  from environmental positions of the H{\sc ii} region. The dotted red
  and black lines mark the peak velocities of Gaussian fits to spectra
  1 and 4, respectively.}
\label{rrl_spectra}
\end{figure}

\subsection{Radio recombination lines}
\label{rrl}

As outlined in Sect. \ref{additions}, for most parts of the survey,
individual RRLs are barely detected, but
after stacking all non-RFI-affected recombination line images in the
velocity domain, we can identify the ionized gas spectral line
emission toward a significant number of sources (28 regions in the
first half of the survey by visual inspection of the
data alone).

To outline the potential of these data, we present a more detailed
look at the data corresponding to the famous H{\sc ii} region M17 (or
G15.06-0.69 in our Galactic nomenclature). Figure \ref{rrl_m17}
shows the zeroth-, first-, and second-moment maps (integrated emission, and
intensity-weighted peak-velocities and line widths) toward the M17
H{\sc ii} region. While the integrated zeroth-moment map clearly shows
the main shell-like H{\sc ii} region surrounding the central OB
cluster (e.g., \citealt{hanson1997,hoffmeister2008}), the interesting
aspect of the RRLs is that we can also study the kinematics of the
ionized gas and set it into context with other components of the
ISM. The right panel of Fig.~\ref{rrl_m17} shows the line-width
distribution, and the broadest lines are found toward
the strongest emission features of the H{\sc ii} region. In addition
to this, Fig. \ref{rrl_m17} (middle panel) clearly shows that the
main arc-like emission of the H{\sc ii} region is associated with gas
peaking around $\sim$5\,km\,s$^{-1}$ , while all the surrounding
ionized gas is shifted to higher velocities around
20\,km\,s$^{-1}$. This shift in peak velocities is also seen in the
individual spectra extracted toward several positions and shown in
Fig.~\ref{rrl_spectra}.

\begin{figure*}[htb] 
\includegraphics[width=0.99\textwidth]{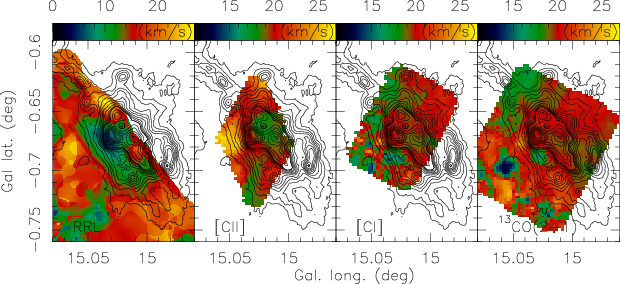}
\caption{Compilation of first moment maps (intensity-weighted peak
  velocities) toward M17 of the THOR RRL data
  (left panel) with ionized carbon, atomic carbon, and molecular carbon
  (second to fourth panel) taken from \citet{perez2012,perez2015}. The
  color scale for the recombination lines is from 0 to
  27\,km\,s$^{-1}$, for the others from 11 to 27\,km\,s$^{-1}$. The
  contours show the 870\,$\mu$m continuum data from the ATLASGAL
  survey starting at a $4\sigma$ level of 200\,mJy\,beam$^{-1}$.}
\label{rrl_m17_compare}
\end{figure*}

The question now arises whether such a velocity shift is only seen in
the ionized gas between the main western emission feature of the H{\sc
  ii} and the environment, or if a similar shift of velocities is
also found in other phases of the ISM. To investigate this in more
detail, Fig.~\ref{rrl_m17_compare} presents the peak velocity maps of
[CII] tracing the weakly ionized gas and the atomic and
molecular carbon components observed in [CI] and $^{13}$CO(2--1) (data
from \citealt{perez2012,perez2015}). The peak
velocities of the molecular and atomic components are both centered on
20\,km\,s$^{-1}$, and the weakly ionized [CII] is found between
15 and 20\,km\,s$^{-1}$. Similar velocity shifts are also found in HI
absorption line studies by \citet{brogan1999} and
\citet{brogan2001}. All environmental gas components therefore
exhibit
velocities shifted by 10 to 15\,km\,s$^{-1}$ relative to the main
velocity found toward the H{\sc ii} region
ridge. \citet{pellegrini2007} modeled the M17 H{\sc ii} region as
being in pressure balance between the radiative and wind components
induced by the central cluster \citep{hanson1997,hoffmeister2008} and
magnetic pressure within the environmental cloud. While clumpiness
also comes into play (e.g., \citealt{stutzki1988}), most features
presented here and in the literature can be reproduced by such
a model (e.g., \citealt{pellegrini2007}).

While individual lines of sight in recombination lines were reported
in the literature from interferometric (e.g.,
\citealt{pellegrini2007}) and single-dish data (e.g.,
\citealt{anderson2011,anderson2015}), full mapping of RRL emission over large samples has been rare (e.g.,
\citealt{urquhart2004}). This is partly caused by the often very weak
recombination line emission and hence inadequate sensitivity. In this
context, THOR now provides an entirely new set of spectrally and
spatially resolved recombination line data toward a large sample of
H{\sc ii} regions because the stacking approach enables us to reach
higher sensitivities than usually possible when observing only single
lines.

Here, we present the data at uniform spatial and spectral resolution,
but in the future, we are planning to re-image individual bright and compact
regions with strong signal at higher spectral resolution. Regions
where the sensitivity for individual lines is adequate can then also
be imaged at higher angular resolution in those individual lines.

\subsection{Polarization and Faraday rotation measures}

Since the polarization calibration and data analysis is far more
complicated than for the rest of the data, we have so far only worked
on the data of the pilot region. The
full survey will be presented and analyzed in a separate paper. Here, we outline
the potential of the survey and highlight initial results toward the
pilot region.  The main data products of the THOR polarization survey
will be a catalog of linearly polarized emission and image cubes of
Stokes $I$, $Q$, and $U$. The catalog will be made by applying
Faraday rotation measure synthesis \citep{brentjens2005}. It will list
fractional polarization and polarization angle at up to three
reference frequencies across the band, and one or more measurements of
the Faraday depth $\phi$, defined through the line-of-sight integral
$$\phi=0.81 \int n_e B_\| dl,$$
where $\phi$ is measured in rad m$^{-2}$, the electron density along
the line of sight $n_e$ in cm$^{-3}$, the component of the magnetic
field projected on the line of sight, $B_\|$ in $\mu$G, and the
line-of-sight distance $l$ in pc (e.g., \citealt{brentjens2005}). A
single line of sight can have more than one Faraday depth depending on
the location of different synchrotron-emitting regions embedded in the
Faraday rotating plasma. The $20\arcsec$ angular resolution of THOR at
the lowest frequency (1 pc at a distance of 10 kpc) reduces confusion
of regions with different Faraday depth inside the synthesized beam,
but true line-of-sight pile-up of emission with different Faraday
depths is always a possibility \citep{brentjens2005}.

\begin{figure}[htb] 
\includegraphics[width=0.49\textwidth]{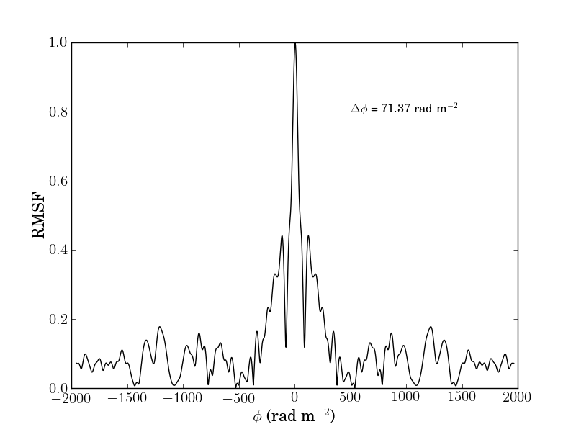}
\caption{RMSF of THOR has a
  central lobe with FWHM $\sim$70\,rad\,m$^{-2}$ with near side lobes at
  the 40\% level around $\pm$200\,rad\,m$^{-2}$. The largest
detectable Faraday
  depth scale is $\sim$130\,rad\,m$^{-2}$.}
\label{RMSF-fig}
\end{figure}

Figure~\ref{RMSF-fig} shows the rotation measure spread function
(RMSF) for the supernova remnant Kes75 in the pilot region. This is
the equivalent of a point spread function in Faraday depth space. The
theoretical resolution in Faraday depth for the THOR survey, following
\citet{brentjens2005}, is 60\,rad\,m$^{-2}$. In practice, rejection of
channels affected by RFI across the band reduces the resolution
somewhat to $\sim 70$\,rad\,m$^{-2}$. More importantly, the loss of
two spectral windows to RFI causes significant broad side lobes at
$\pm$ 300\,rad\,m$^{-2}$ that can interfere with the detection of
faint components. The highest Faraday depth that can be detected by
THOR before Faraday rotation across a single frequency channel
depolarizes the emission is $1.6 \times 10^4$
\citep{brentjens2005}. The Faraday depth resolution of THOR opens a
new part of parameter space in terms of exploring structure in Faraday
depth and its ability to detect very large Faraday depths.

The widest continuous range in Faraday depth that can be recognized
in THOR is given by the shortest wavelength covered
\citep{brentjens2005}. The simplest form of a continuous Faraday depth
structure is a region filled with plasma that also emits synchrotron
emission everywhere along the line of sight. Depending on the shortest
wavelength observed, there is a largest Faraday depth scale that can
be recognized, similar to the missing short spacing problem in radio
interferometry. For THOR, the largest extent in Faraday depth is
133\,rad\,m$^{-2}$, yielding a dynamic range for Faraday depth
structure of a factor $\sim 2$. To set this into perspective, a region
with a size of $\gtrsim 100$\,pc, $n_e \sim 0.1$\,cm$^{-3}$, $B \sim
5$\,$\mu$G and $\Delta \phi/\phi \sim 1$ would produce detectable
Faraday complexity in THOR. There are many different permutations of
line of sight depth, electron density, and magnetic field strength
that can produce resolved structure in Faraday depth.
 
\begin{figure}[htb] 
\includegraphics[width=0.49\textwidth]{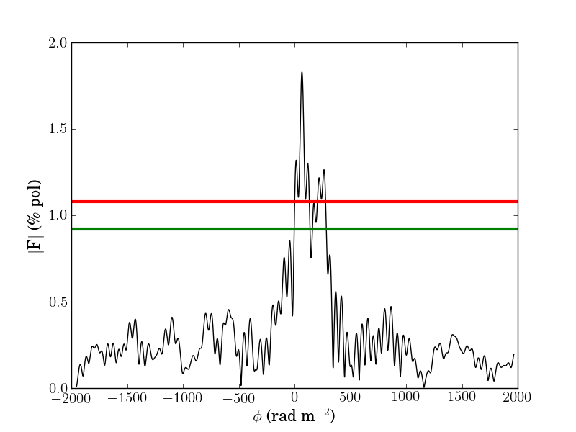}
\caption{Integrated Faraday depth spectrum of the pulsar wind nebula
  at the center of the SNR Kes75 ($l=29.7104^{\circ}$ \&
  $b=-0.2402^{\circ}$). The vertical axis represents the degree of
  polarization as a percentage of the total flux density, after
  unwrapping Faraday rotation assuming Faraday depth $\phi$ and
  averaging over the observed frequency range. The green and
  red lines indicate the maximum over all Faraday depths of 100 and 300
  realizations of the analysis, respectively, which replace the $Q$ and
  $U$ spectra of the target by noise spectra from off-source
  positions, integrated over the same solid angle as the target. We
  detect components at Faraday depth $\phi_1 = 60$ rad m$^{-2}$ and
  Faraday depth $\phi_2 = 258$ rad m$^{-2}$. The formal error in
  Faraday depth is $\sim$5 rad m$^{-2}$ for a 10$\sigma$ detection.}
\label{kes75-fig}
\end{figure}

Figure~\ref{kes75-fig} demonstrates how these numbers combine for the
SNR Kes 75. This figure shows the result of Faraday
rotation measure synthesis using Stokes $Q$ and $U$ spectra integrated
over the pulsar wind nebula. After division by Stokes $I$, the
amplitude of the Faraday depth spectrum is expressed as a percentage
of the total flux density. We define the detection statistics by
repeating the analysis 300 times with the Stokes $Q$ and $U$ replaced
with equivalent spectra extracted from empty regions in the image. The
highest noise peak is used in Fig.~\ref{kes75-fig} as the detection
threshold for the on-source Faraday depth spectrum. We also apply a
lower limit of 1\% for detected polarized emission to avoid false
detections related to residual instrumental polarization.

We find two peaks that exceed the 1-in-300 detection threshold. The
peaks are separated by 200 rad m$^{-2}$, which is well beyond the
Faraday depth resolution of the data. This is a very broad Faraday
depth range for the angular scale $< 1\arcmin$. The separation between
the peaks is more than the largest Faraday depth scale that can be
observed in THOR. We can therefore not exclude an additional broad
component with a Faraday depth scale comparable to the separation of
the peaks. Repeating the analysis by separating the band in thirds by
$\lambda^2$ indicates a gradual decrease in fractional polarization
from 3.3\% at the upper frequency range to 1.5\% at the center, and no
formal detection in the lowest frequency spectral window. The THOR
polarization catalog will produce this level of information for all
detected sources. By contrast, different regions of the bright shell
of Kes 75 only show peaks at Faraday depths near 0 with amplitudes
$\lesssim 1\%$ that are consistent with residual polarization leakage.

THOR will provide polarized background sources for studies of the
Galactic magnetic field on scales spanning four orders of magnitude, as
well as measurements of Faraday depth structure for individual objects
in the Milky Way. The angular resolution and polarization information
will make this survey ideal for magnetic field structure and
depolarization in supernova remnants, and for the detection of young supernova
remnants embedded in crowded star formation regions.

\section{Discussion}

Based on the early results presented in the previous section, we
envision a multitude of future scientific applications. The advantage
of THOR is that we do not have to rely on single case studies but that
larger statistical approaches are possible. For example, the HI study
of the W43 cloud presented in \citet{bihr2015b} will be extended to
many clouds within the Milky Way. Similarly, feedback studies as
indicated by the M17 data in Sect. \ref{rrl} will be extended to the
whole sample of detected H{\sc ii} region in the radio recombination
line emission. In a different application, the HI data enable
studying the density fluctuation structure function of the CNM down to a
few tens of arcsec scales. This structure function is related to the
ISM turbulence and can be directly compared to different theoretical
models to constrain turbulence and energy dissipation mechanisms
(e.g., \citealt{dickey2001,audit2010}).

For OH masers and absorption studies, THOR provides the
perfect dataset. However, for thermal OH emission, the extended
structures are filtered out by our C-configuration observations. We are
currently exploring whether complementing the THOR OH data with short
spacing from the SPLASH survey \citep{dawson2014} or complementary
Effelsberg/GBT observations is sufficient, or if the shorter
baselines from the VLA in D-configuration are needed for such an aspect of the
ISM studies. Similarly, the continuum data allow us to derive spectral
indices for compact structures \citep{bihr2016}, but spectral
indices for more extended sources such as SNRs are much
harder to determine from THOR data alone. Therefore, we are currently
examining whether single-dish short spacings are sufficient for the
science goals related to the continuum emission in the survey, or
if D-configuration data may be needed.

We currently merely scratch the surface of the polarization aspect of
the THOR survey. Since we have observed the full polarization, Faraday
rotation and magnetic field studies of the Milky Way will be
feasible. However, the data calibration, imaging, and analysis aspect
of this part of the survey have yet to be realized, and therefore,
polarization and magnetic field studies will be presented in
forthcoming publications (e.g., Stil et al.~in prep.).

In addition to the THOR data as a stand-alone survey, it will obviously
be important to combine THOR with existing surveys at other
wavelengths. Only then we will be able to address all facets of the
Milky Way in its appropriate depth. Understanding Galactically
important regions such as the bar--spiral arm interface can directly be
compared with extragalactic studies (e.g., THINGS,
\citealt{walter2008}) and thus be set into a global context.  The
combination of Galactic and extragalactic systems allows us to derive
a concise and complete picture of the ISM and star formation
processes.

Furthermore, THOR can also be considered as a precursor of Square
Kilometer Array (SKA) pathfinder science because the planed GASKAP
survey (The Galactic ASKAP survey) with the Australian SKA Pathfinder
telescope will achieve comparable sensitivities and angular resolution
elements in the southern hemisphere \citep{dickey2013}.

\section{Summary}
\label{conclusion}

We presented the survey specifications, scientific goals, and early
results of the new Galactic plane survey THOR: The
HI/OH/Recombination line survey of the Milky Way.  We release the data
stepwise, including the first half of the data in this paper. The
remaining data will be provided successively after the ongoing
calibration/imaging process. The data can be accessed at the project
web-page at http://www.mpia.de/thor.

THOR observes the spectral lines of HI, OH, and several radio
recombination lines as well as the continuum emission from 1 to 2\,GHz
in full polarization over approximately 132 square degrees between
Galactic longitudes of 14.5 and 67.4\,deg and latitudes $\pm
1.25$\,deg. These data allow us to study the different phases of the
ISM from the atomic HI to the molecular OH and the ionized gas in the
recombination and continuum emission. This enables studies of the
atomic to molecular gas conversion, molecular cloud formation,
feedback processes from the forming H{\sc ii} regions, and
magnetic field studies of the ISM. We showed selected results from these
datasets. In addition to using THOR by itself, it will also be useful
in conjunction with many other existing Galactic plane surveys to
study the interplay of the various components of our Milky Way.

\begin{acknowledgements} 
  The National Radio Astronomy Observatory is a facility of the
  National Science Foundation operated under cooperative agreement by
  Associated Universities, Inc. We like to thank Juan Pablo
  Perez-Beaupuits for providing the ionized, atomic, and molecular data
  of M17 presented in Fig.~\ref{rrl_m17_compare}.
\end{acknowledgements}

%\bibliography{/home/beuther/tex/bibliography}   
%\bibliography{/Users/henrikbeuther/tex/bibliography}
%\bibliographystyle{aa}    % this does the style, aa.bst necessary
%\input{thor_0816_v2.bbl}

\clearpage 

\appendix

\section{Noise maps}
\label{noise_maps}

\begin{figure*}[htb] 
\includegraphics[width=0.99\textwidth]{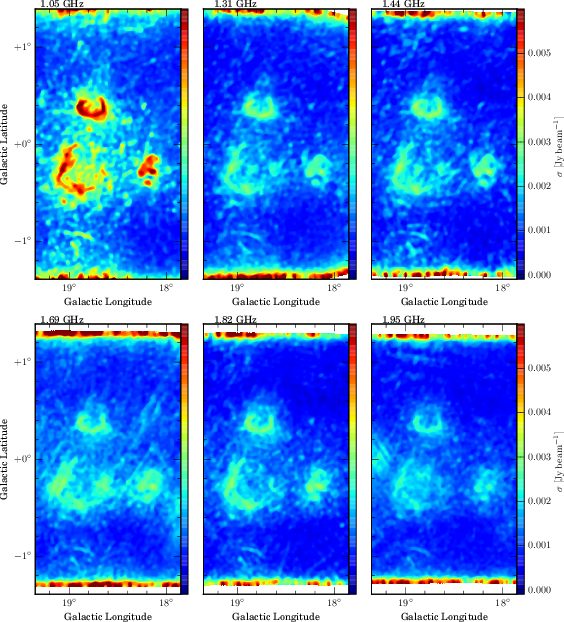}
\caption{Example continuum noise maps in the l 18 to 19\,deg
  field. The scale bars for all wavelengths are the same. The
  frequencies increase from the top left to the bottom right
corner. The
  noise level clearly depends on the source structure and the
  different bands.}
\label{noise_cont}
\end{figure*}

\begin{figure*}[htb] 
\includegraphics[width=0.99\textwidth]{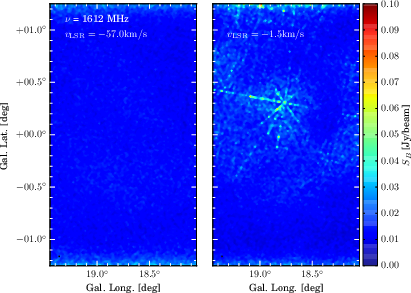}
\caption{Example noise maps in the l 18 to 19\,deg field for OH at
  1612\,MHz. The left panel shows an empty channel, while the right
  panel shows an example for a channel with a strong maser peak. The
  scale bars for all wavelengths are the same.}
\label{noise_oh}
\end{figure*}

\end{document}